\pdfoutput=1
\documentclass[a4paper,onecolumn]{article}
\usepackage[top=2.5cm,bottom=2.5cm,left=2.5cm,right=2.5cm]{geometry}  
\usepackage{authblk}
\usepackage{bbm}
\usepackage{braket}
\usepackage{graphicx}
\usepackage{dsfont}
\usepackage[T1]{fontenc}
\usepackage[utf8]{inputenc}
\usepackage{float}
\usepackage{braket}
\usepackage{indentfirst}
\usepackage{enumerate}
\usepackage{url}
\usepackage[dvipsnames,x11names]{xcolor} 	
\usepackage[stable]{footmisc}
\usepackage{fancyhdr}
\usepackage[all]{xy}
\usepackage{graphicx}
\usepackage{tikz}
\usetikzlibrary{calc,positioning,arrows.meta, shapes, positioning}
\usepackage{framed}
\usepackage{ytableau}
\usepackage{youngtab}
\usepackage{mathtools}
\usepackage{enumerate}
\usepackage[title]{appendix}
\usepackage{textcomp}
\usepackage{newcent}
\usepackage[sc,noBBpl]{mathpazo}
\usepackage{amsmath,amsfonts,amssymb,amsthm}
\usepackage[mathscr]{eucal}
\usepackage{tcolorbox}
\usepackage[colorlinks=true,citecolor=PineGreen,linkcolor=RoyalBlue,urlcolor=Violet]{hyperref}
\usepackage{caption} 
\usepackage{subcaption}
\usepackage[
backend=biber,
style=nature,
sorting=none
]{biblatex}
\AtEveryBibitem{\clearfield{month}}
\AtEveryBibitem{\clearfield{day}}
\AtEveryBibitem{\clearfield{issue}}

\usepackage{algorithm}
\usepackage{algpseudocode}  
\usepackage{float}

\newcommand{\tr}{\operatorname{tr}}
\theoremstyle{plain}
\newtheorem{theorem}{Theorem}
\newtheorem{lemma}[theorem]{Lemma}
\newtheorem{proposition}[theorem]{Proposition}
\newtheorem{corollary}[theorem]{Corollary}

\newtheorem{definition}[theorem]{Definition}
\newtheorem{fact}[theorem]{Fact}

\newtheoremstyle{note}{\topsep}{\topsep}{\slshape}{}{\scshape}{}{ }{}
\theoremstyle{note}

\newtheorem{remark}[theorem]{Remark}

\newcommand{\id}{\mathds{1}}

\newcommand{\<}{\langle}
\renewcommand{\>}{\rangle}

\newcommand\be{\begin{equation}}
\newcommand\ee{\end{equation}}
\newcommand\bea{\begin{array}}
	\newcommand\eea{\end{array}}
\newcommand\ben{\begin{eqnarray}}
\newcommand\een{\end{eqnarray}}

\newcommand\bei{\begin{itemize}}
	\newcommand\eei{\end{itemize}}
\newcommand\bee{\begin{enumerate}}
	\newcommand\eee{\end{enumerate}}

          \newcommand\hlight[1]{\tikz[overlay, remember picture,baseline=-\the\dimexpr\fontdimen22\textfont2\relax]\node[rectangle,fill=white!50,rounded corners,fill opacity = 0.2,draw,thick,text opacity =1] {$#1$};}

\newcommand{\h}{\mathcal{H}}

\usepackage{adjustbox}
\usetikzlibrary{decorations.pathreplacing}

\begin{document}
\title{Iterative construction of $\mathfrak{S}_p \times \mathfrak{S}_p$ group-adapted irreducible matrix units for the walled Brauer algebra }
\author[1]{Michał Horodecki}
\author[1]{Michał Studziński\thanks{Corresponding author: michal.studzinski@ug.edu.pl}}
\author[2]{Marek Mozrzymas}
\affil[1]{\small{\textit{International Centre for Theory of Quantum Technologies, University of Gdańsk, Jana Bażyńskiego 1A, 80-309 Gdańsk, Poland}}}
\affil[2]{\small{\textit{Institute for Theoretical Physics, University of Wrocław, plac Maxa Borna 9, 50-204 Wrocław, Poland}}}

\date{}
\maketitle
\begin{abstract} 
In this work, we present an algorithmic treatment of the representation theory of the algebra of partially transposed permutation operators, denoted by $\mathcal{A}^d_{p,p}$, which is a matrix representation of the abstract walled Brauer algebra.

We provide an explicit and fully developed framework for constructing irreducible matrix units within the algebra. In contrast to the established earlier Gelfand-Tsetlin type constructions, the presented matrix units are adapted to the action of the subalgebra $\mathbb{C}[\mathfrak{S}_p] \times \mathbb{C}[\mathfrak{S}_p]$, where $\mathfrak{S}_p$ is the symmetric group. What is more, the basis is constructed in such a way that it produces the decomposition of the algebra into a direct sum of ideals, in contrast to its nested structure considered before. The decomposition of this kind has not been considered before in full generality. Our method reveals a recursive scheme for generating irreducible matrix units in all ideals of $\mathcal{A}^d_{p,p}$, offering a systematic approach that applies to small system sizes and arbitrary local dimensions. We apply the developed formalism to the algebra $\mathcal{A}^d_{2,2}$ and illustrate the algorithm in practice. In addition, using the constructed basis, we proved a novel contraction theorem for the elements from  $\mathcal{A}^d_{3,3}$, which is the starting point for further investigations.
\end{abstract}

\section{Dedication}
{\it Blessed be the chaos, 

for out of it form will come. 

\hskip 3mm

Blessed be the light, 

for it will divide us 

from darkness.}

\hskip 3mm

\noindent
(from "Still life", R. Horodecki,  poetry book,
{\it Sum ergo Cogito},
translated by  J. Ward)

\hskip 3mm

It was about fifteen years ago, when two of us (MH and MS) were taking our first unsteady steps in representation theory.
We needed it in order to compute the entanglement of a certain class of states. And then Ryszard Horodecki said: tell Marek Mozrzymas about your problem. He was just hosting Marek at that time in Sopot, in a beautiful villa at Andersa Street, which for many years had been the seat of the National Quantum Information Centre. And that was when it began. We formed a team, and over these all years of collaboration, the chaos that we had in our minds was gradually transformed into form. We believe that together with our collaborators we have shed a bit of light on the objects invariant with respect to unitaries of the form $U^{\otimes n}\otimes \overline{U}^{\otimes n}$, 
the topic to which also this paper is devoted.

\section{Introduction}
In this work, we address the construction of irreducible matrix representations of the walled Brauer algebra~\cite{Bra37} that are group-adapted—that is, representations in which the action of the algebra is explicitly compatible with a sequence of subalgebra restrictions, particularly those induced by the natural action of the product algebra $\mathbb{C}[\mathfrak{S}_p] \times \mathbb{C}[\mathfrak{S}_p]$, where $\mathfrak{S}_p$ is the symmetric group. Such compatibility can be useful for applications involving partial symmetries, such as those encountered in multipartite quantum systems~\cite{Koch_2012,grinko2023linearprog,StudzinskiIEEE22,grinko2025phd} and invariant theory~\cite{10.1063/1.1665059,10.1063/1.1665778,Halverson1996CharactersOT,KOIKE198957}.

Existing constructions of matrix elements for the walled Brauer algebra are typically generalizations of the classical Young–Yamanouchi basis~\cite{young1931quantitative, yamanouchi1937construction} and branching graph techniques~\cite{ceccherini2010representation}. While these approaches yield correct and explicit representations, they generally fail to exhibit group-adaptedness in the above sense, and do not align naturally with the subalgebra structure dictated by $\mathbb{C}[\mathfrak{S}_p] \times \mathbb{C}[\mathfrak{S}_p]$. 

We present a systematic algorithm that generates irreducible matrix units in a way that respects the hierarchical subalgebra structure in the semisimple image of the abstract algebra. Our method provides an explicit and canonical procedure for building these representations, which can be directly implemented for both symbolic and numerical computations. Furthermore, the algorithm sheds new light on the internal combinatorial structure of the walled Brauer algebra, and its compatibility with group-theoretic frameworks, including Schur–Weyl duality~\cite{fulton_harris,Goodman,harrow2005} and its generalizations to the mixed Schur-Weyl duality~\cite{KOIKE198957,BEN96}.

While the walled Brauer algebra is defined as an abstract algebra, see Figure~\ref{fig:WBA_abstract} below, its applications - to entanglement theory~\cite{Collins2024teleportationofpost,Huber2022ER,Yongzhang2021permutation,PhysRevA.63.042111} or to various quantum informational processing tasks and methods~\cite{harrow2005,grinko2025phd,grinko2024efficientquantumcircuitsportbased,fei2023efficientquantumalgorithmportbased,PRXQuantum.5.030354,nguyen2023mixedschurtransformefficient,marvian2021qudit,marcinska24,StudzinskiIEEE22,PhysRevA.98.063815}
- require concrete representations on tensor product spaces. In particular, the algebra manifests in physical models through the so-called \emph{the algebra of partially transposed permutation operators} $\mathcal{A}^d_{p,p}$~\cite{KOIKE198957,BENKART1994529,Moz1,MozJPA,grinko2023gelfandtsetlinbasispartiallytransposed}, which act on spaces of the form \((\mathbb{C}^d)^{\otimes 2p}\) and generate a finite-dimensional, semi-simple image of the abstract algebra. This image plays a key role in analysing the properties and internal structure of quantum protocols. The structure of the map giving the semi-simple image is quite complicated: the map has a non-trivial kernel, and its structure is not yet easy to phrase~\cite{ANDERSEN_STROPPEL_TUBBENHAUER_2017}.  Nevertheless, the conditions for mapping abstract algebra objects to their matrix representations do not require full knowledge about the kernel's structure and do not affect our results.

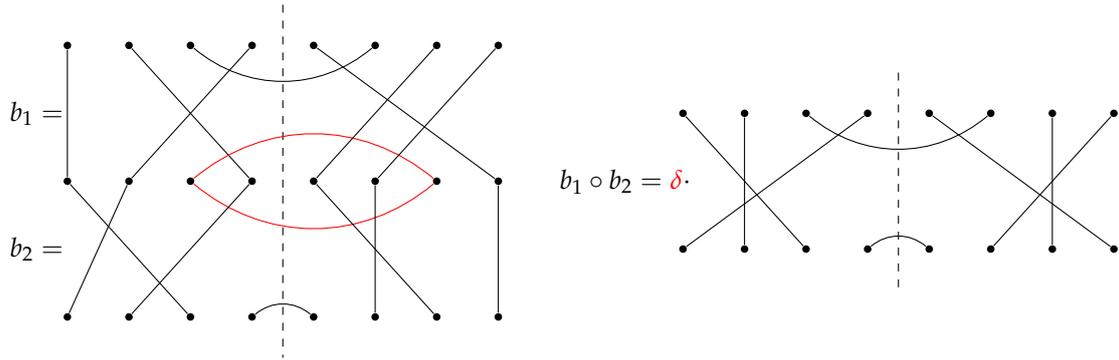
\begin{figure}[h!]
\centering
\begin{tikzpicture}[scale=0.9, every node/.style={inner sep=2pt}]
  \def\n{8}
  \def\sep{0.9}
  \def\h{2.0}

  \foreach \i in {1,...,8} {
    \node[circle, fill=black, inner sep=1pt] (s\i) at ({\i*\sep}, {2*\h}) {};
    \node[circle, fill=black, inner sep=1pt] (sp\i) at ({\i*\sep}, {\h}) {};
  }

  \foreach \i in {1,...,8} {
    \node[circle, fill=black, inner sep=1pt] (pp\i) at ({\i*\sep}, {0}) {};
  }

  \node at (0.5*\sep, 1.5*\h) {$
    b_1 =$};
  \node at (0.5*\sep, 0.5*\h) {$
    b_2 =$};

  \draw[dashed] (4.5*\sep, {2.3*\h}) -- (4.5*\sep, {-0.3*\h});

  \draw (s1) -- (sp1);
  \draw (s2) -- (sp4);
  \draw (s3) to[bend right=40] (s6);
  \draw[red] (sp3) to[bend left=40] (sp7);
  \draw (s4) -- (sp2);
  \draw (s5) -- (sp8);
  \draw (s7) -- (sp5);
  \draw (s8) -- (sp6);

  \draw (sp1) -- (pp3);
  \draw (sp2) -- (pp1);
  \draw[red] (sp3) to[bend right=40] (sp7); 
  \draw (sp4) -- (pp2);
  \draw (pp4) to[bend left=40] (pp5);
  \draw (sp5) -- (pp7);
  \draw (sp6) -- (pp6);
  \draw (sp8) -- (pp8);

  \def\compShift{-1.0}
  
  \foreach \i in {1,...,8} {
    \node[circle, fill=black, inner sep=1pt] (c\i) at ({(\i+10)*\sep}, {2*\h+ \compShift}) {};
    \node[circle, fill=black, inner sep=1pt] (cp\i) at ({(\i+10)*\sep}, {\h+ \compShift}) {};
  }

  \draw[dashed] (14.5*\sep, {2.3*\h+ \compShift}) -- (14.5*\sep, {0.7*\h+ \compShift});
  \node at (10.5*\sep, 1.5*\h+ \compShift) {$
    b_1 \circ b_2 =\textcolor{red}{\delta} \cdot \qquad$};

  \draw (c1) -- (cp3);
  \draw (c2) -- (cp2);
  \draw (c3) to[bend right=40] (c6);
  \draw (c4) -- (cp1);
  \draw (c5) -- (cp8);
  \draw (c7) -- (cp7);
  \draw (c8) -- (cp6);
  \draw (cp4) to[bend left=40] (cp5);

\end{tikzpicture}
\caption{\small 
The \emph{walled Brauer algebra} \(\mathcal{B}^{\delta}_{m,n}\), where \(m,n \geq 0\), and \(\delta \in \mathbb{C}\), was introduced in a series of  works including~\cite{Bra37,VGTuraev_1990,KOIKE198957,BENKART1994529,BEN96,bulgakova:tel-02554375,Cox1}. This algebra is defined abstractly as a linear span over \(\mathbb{C}\) of certain diagrammatic elements. Each diagram consists of two rows of \(m + n\) nodes, separated by a vertical wall placed between the first \(m\) and the remaining \(n\) nodes. Edges connect nodes in pairs, subject to the following rules: (a) Pairs of nodes within the same row must lie on opposite sides of the wall. (b) Pairs of nodes between different rows must lie on the same side of the wall.
This combinatorial structure defines the multiplication in the algebra via concatenation of diagrams, with closed loops contributing a scalar factor of \(\delta\). The dimension of \(\mathcal{B}^{\delta}_{m,n}\) coincides with the number of such valid pairings, and in particular equals \((m+n)!\), matching the size of the symmetric group \(\mathfrak{S}_{m+n}\). Here we present an example of graphical composition of two diagrams \(b_1,b_2 \in \mathcal{B}_{4,4}^\delta\). Identifying a closed loop (in red) results in multiplying the diagram by a scalar \(\delta \in \mathbb{C}\). We see that the composition \(b_1 \circ b_2\) remains within \(\mathcal{B}_{4,4}^\delta\).}
\label{fig:WBA_abstract}
\end{figure}

The paper is organised as follows:
\begin{itemize}
    \item In Section~\ref{sec:tech_intro} we introduce basic notations and fact for the group algebra $\mathbb{C}[\mathfrak{S}_p]$ and the algebra of the partially transposed permutation operators $\mathcal{A}^d_{p,p}$. In principle, we recall Lemma~\ref{lemma1} from another paper, establishing an important fact on composing a special class of elements from the $\mathcal{A}^d_{p,p}$.
    \item In Section~\ref{sec:comparison} we approach the problem of constructing group-adapted irreducible bases in $\mathcal{A}^d_{p,p}$ from the perspective of the existing literature. We carefully summarize the existing knowledge on the topic and present a rough idea of attacking the problem in the most general situation. 
    \item In Section~\ref{sec:algo} we present in detail an algorithm for constructing irreducible matrix units in $\mathcal{A}^d_{p,p}$ which are group-adapted with respect to $\mathbb{C}[\mathfrak{S}_p] \times \mathbb{C}[\mathfrak{S}_p]$ producing a decomposition of the algebra into a direct sum of ideals. Until now, existing methods for constructing irreducible matrix units have not taken the group-adapted property into account~\cite{grinko2023gelfandtsetlinbasispartiallytransposed}. The only exceptions are the two highest ideals of the algebra $\mathcal{A}^d_{p,p}$, where such a group-adapted basis has been constructed~\cite{StudzinskiIEEE22,studziński2025WBA}. Here, we significantly extend and unify the previous approaches.
    \item In Section~\ref{sec:decomposition}, for technical reasons, we introduce a new set of orthogonal operators producing the linear span of the algebra $\mathcal{A}_{p,p}^d$. These operators, contrary to previous constructions~\cite {studziński2025WBA, grinko2023gelfandtsetlinbasispartiallytransposed}, are always given as a linear combination of elements from two neighbouring ideals of the algebra. Having this set, we formulate a series of results implied by its properties. In particular, we show how these results yield a decomposition into a direct sum of the two highest ideals, and how they can be used to construct group-adapted irreducible matrix units within the considered ideals. 
    \item In Section~\ref{sec:Example2v2} we present how our algorithm works in practice. We focus on the case of $\mathcal{A}^d_{2,2}$. We present the full set of group-adapted irreducible matrix units for $\mathcal{A}^d_{2,2}$ and present its decomposition into a direct sum of ideals, each isomorphic with a certain matrix complex algebra.
    \item We start Section~\ref{sec:Example3v3} by formulating general results concerning the algebra $\mathcal{A}_{p,p}^d$ for any $p\geq 1$.  Then we apply these results and matrix units derived in Section~\ref{sec:Example2v2} to formulate a novel contraction theorem for the elements from $\mathbb{C}[\mathfrak{S}_3] \times \mathbb{C}[\mathfrak{S}_3]$ sandwiched by a single arc operator from $\mathcal{A}^d_{3,3}$. Such contraction theorems are necessary in proving further various properties of the considered algebra, in particular in constructing irreducible matrix units.
    \item The paper ends with a discussion Section~\ref{sec:disscus} where we summarize our findings and present potential further research directions.
\end{itemize}

\section{Basic facts and notations used throughout the manuscript}
\label{sec:tech_intro}
Before proceeding with the main content of this paper, we introduce the key definitions and notations that will be used throughout this section and the remainder of the manuscript. For more details, we refer to the standard textbooks~\cite{fulton1997young, georgi2000lie, ceccherini2010representation}. 

A partition $\mu$ of $p$, denoted as $\mu \vdash p$, is a non-decreasing sequence of positive integers $(\mu_1, \ldots, \mu_r)$ for any non-negative integer $r$ with $\mu_1\geq \cdots \geq \mu_r > 0$ which sums up to $\sum_{i=1}^{r} \mu_i = p$.
A partition $\mu \vdash p$ can be represented as a Young diagram with $p$ boxes, which has $\mu_i$ boxes in the $i$-th row.
For instance, $\mu=(3,1)$ can be represented as a Young diagram given by
\begin{align}
\label{eq:example_young_diagram}
    \mu = \ydiagram{3,1}.
\end{align}
On Young diagrams, we are allowed to perform operations of adding/subtracting a box $\Box$. We denote these operations as $\mu+\Box, \mu-\Box$ respectively. For instance, for the Young diagram $\mu$ given in Eq.~\eqref{eq:example_young_diagram}, $\mu+\square$ and $\mu-\square$ are given by
\begin{align}
    \alpha+\square \in \left\{ \ydiagram{4,1}, \ydiagram{3,2}, \ydiagram{3,1,1} \right\},
    \quad
    \alpha-\square \in \left\{ \ydiagram{2,1}, \ydiagram{3} \right\}.
\end{align}
For a given $p$, in the set of all Young diagrams, we introduce the relation $\mu \sim _{\square }\nu $ meaning that the corresponding Young
diagrams $\mu,\nu \vdash p$ differ by a box $\Box$. What is more, we have the following property
\begin{equation}
\mu \sim _{\square }\nu \Longleftrightarrow \exists !\tau \vdash (p-1)
:\quad \tau = \mu-\Box \quad \wedge \quad \tau =\nu-\Box.
\end{equation}

There is a well-known relation between the number of Young diagrams for fixed $p$ and irreducible representations (irreps) of the $\mathfrak{S}_p$. Namely, the number of all inequivalent irreps of $\mathfrak{S}_p$ equals the number of all Young diagrams. In other words, every Young diagram $\mu \vdash p$ labels an irrep of $\mathfrak{S}_p$. The set of all possible irreps of $\mathfrak{S}_p$, equivalently, the set of all Young diagrams $\mu \vdash p$ will be denoted as $\widehat{\mathfrak{S}}_p$. Whenever it is clear from the context, we use the  symbol $\mu \vdash p$ to denote a Young diagram or the respective irrep of $\mathfrak{S}_p$.

For all $\sigma \in \mathfrak{S}_p$ define their natural representation $V:\mathfrak{S}_p \rightarrow  (\mathbb{C}^d)^{\otimes p}$ as
\begin{align} \label{eq:actionS_k} 
V_\sigma (\ket{v_1}\otimes \ket{v_2} \otimes \dots \otimes \ket{v_p}) := \ket{v_{\sigma^{-1}(1)}}\otimes \ket{v_{\sigma^{-1}(2)}} \otimes \dots \otimes \ket{v_{\sigma^{-1}(p)}}.
\end{align}
This representation naturally extends to the representation of the group algebra $\mathbb{C}[\mathfrak{S}_p]=\operatorname{span}_{\mathbb{C}}\{V_\sigma \ : \ \sigma \in \mathfrak{S}_p\}$. On the space $(\mathbb{C}^d)^{\otimes p}$ every irreducible component $\mu$ of $\mathbb{C}[\mathfrak{S}_p]$ is spanned by irreducible matrix units given as~\cite{Ram1992MatrixUF} 
\begin{align} \label{eqn:basis_Eij}
E_{ij}^{\mu} = \frac{d_\mu}{p!} \sum_{\sigma \in \mathfrak{S}_p} \varphi_{ji}^\mu (\sigma^{-1}) V_\sigma,\qquad E_{ij}^\mu E_{kl}^\nu = \delta^{\mu \nu} \delta_{jk} E_{il}^{\mu}, \qquad \tr(E_{ij}^\mu) = m_{\mu} \delta_{ij}.
\end{align} 
where $m_\mu$ is the multiplicity of the irrep $\mu$, $\varphi_{ji}^\mu(\sigma^{-1})$ are the matrix elements of the irrep $\mu \vdash p$ of  $\sigma^{-1} \in \mathfrak{S}_p$, and $i,j = 1, \dots, d_\mu$, with $d_\mu$ being the dimension of the irrep $\mu$. The numbers $m_\mu, d_\mu$ can be evaluated by the combinatorial formulas, called the hook-content and the hook-length formula, respectively~\cite{fulton1997young, georgi2000lie, ceccherini2010representation}.  The operators $E_{ij}^{\mu}\in \operatorname{End}[(\mathbb{C}^d)^{\otimes 2p}]$ are non-zero operators if and only if the height $h(\mu)$ of the Young diagram $\mu$ satisfies $h(\mu)\leq d$. Throughout the manuscript, we implicitly assume this condition when working with $E_{ij}^\mu\in \operatorname{End}[(\mathbb{C}^d)^{\otimes 2p}]$, unless stated otherwise.
Using equations~\eqref{eqn:basis_Eij} one can write an arbitrary permutation operator $V_\sigma$ as a linear combination of the operators from~\eqref{eqn:basis_Eij}
\begin{align}
\label{eq:VasE}
\forall \ \sigma\in \mathfrak{S}_p \qquad V_\sigma=\sum_{\mu}\sum_{ij}\varphi^{\mu}_{ij}(\sigma)E^{\mu}_{ij}.
\end{align}
This equation together with ~\eqref{eqn:basis_Eij} allows us to write explicitly the left and the right action of an arbitrary operator $V_\sigma$ on irreducible matrix units~\eqref{eqn:basis_Eij}:
\begin{align}
\label{eq:actionVonE}
V_\sigma E_{ij}^\mu V_{\sigma^{-1}}=\sum_{k,l}\varphi^{\mu}_{ki}(\sigma)\varphi^{\mu}_{jl}(\sigma^{-1})E_{kl}^{\mu}.
\end{align}
To construct the irreducible matrix elements $\varphi_{ji}^\mu(\sigma)$ (so in fact the irreducible matrix units), one can use, for example, the Young-Yamanouchi construction~\cite{young1931quantitative, yamanouchi1937construction}. The Young-Yamanouchi construction is a subgroup-adapted scheme, and it relies on the chain of inclusions
\begin{align}
\label{S_n:inclusions}
\mathfrak{S}_1\subset \mathfrak{S}_2 \subset \cdots \subset \mathfrak{S}_p
\end{align}
which is multiplicity-free~\cite{Okunkov,Vershik2005,ceccherini2010representation}. In other words, to obtain irreps for the group $\mathfrak{S}_k$, it is enough to know the irreps for $\mathfrak{S}_{k-1}$.  This process is memory-dependent, meaning we have to keep track of the whole path on the chain when constructing irreps on the higher layers. To distinguish between two different directions of the construction, we introduce an additional notation. For example, when having $p$ systems, we construct operators $E^\mu_{ij}$ by considering the chain~\eqref{S_n:inclusions} starting from the $1^{\text{st}}$ system on the left or from the $p^{\text{th}}$ system on the right. Then, to underline the direction, we use arrows and write $\overrightarrow{E}^\mu_{ij}$. When, however, constructing the basis in the opposite order, we write $\overleftarrow{E}^\mu_{ij}$. This ordering is strictly related to the irreps induction process from $\mathfrak{S}_{k-1}$ to $\mathfrak{S}_{k}$. The mentioned direction of the irrep construction is important later in the manuscript and allows us to perform elegant calculations of the partial traces with respect to the last system in the chosen order~\cite{Ram1992MatrixUF,StudzinskiIEEE22}. Namely, the irreducible matrix unit of $\mathbb{C}[\mathfrak{S}_{k}]$, partially traced over the last system, is again an irreducible matrix unit, but this time of $\mathbb{C}[\mathfrak{S}_{k-1}]$. This process can be applied iteratively. To perform such a derivation, we need to keep track of the basis construction from $\mathbb{C}[\mathfrak{S}_{k-1}]$ to $\mathbb{C}[\mathfrak{S}_{k}]$. In other words, for a given $\mu \vdash k$, we need to know from which $\alpha \vdash (k-1)$ it can be obtained. Then, for fixed $\mu \vdash k$, every index $i\equiv i_\mu$ can be described by a unique pair of $\alpha \vdash (k-1)$ and the index  $i_\alpha$ describing the position within the irrep $\alpha$. Then every irreducible matrix unit $\overleftarrow{E}^\mu_{ij} \in \mathbb{C}[\mathfrak{S}_{k}]$ can be written in terms of irreducible basis adapted to $\mathbb{C}[\mathfrak{S}_{k-1}]$ as:
\begin{align}
\label{eq:notationPRIR1}
E^{\mu}_{ij}\equiv E^{\mu}_{I_\alpha J_{\alpha'}},
\end{align}
where 
\begin{equation}
\label{eq:notationPRIR2}
I_{\alpha }=\left( 
\begin{array}{c}
\alpha \\ 
i_{\alpha }
\end{array}
\right) ,\quad J_{\alpha ^{\prime }}=\left( 
\begin{array}{c}
\alpha ^{\prime } \\ 
j_{\alpha ^{\prime }}
\end{array}
\right).
\end{equation}

In the part of the manuscript dedicated purely to the algorithm, we can use even simpler notation. Namely, all objects of the form $\overrightarrow{E}^\mu_{ij} \otimes \overleftarrow{E}^\nu_{rs} \in \mathbb{C}[\mathfrak{S}_k] \times \mathbb{C}[\mathfrak{S}_k]$, where $1\leq k\leq p$, for simplicity will be written as
\begin{align}
\label{eq:simplifiedE}
\overrightarrow{E}^{k}_L\otimes \overleftarrow{E}^{k}_R,   
\end{align}
where the lower indices $L,R$ contain all the required information.
It means we are interested only in the number of systems involved, the side of the wall on which a given operator acts (indices $L$ and $R$), and the direction of the basis construction. Equivalently, our focus will be solely on the classification of the operator within a specific structural class of objects.

In a bipartite setting, the partial transposition \(t_1\) with respect to the first system is expressed as:
\begin{align}
t_1:\; |v_i\rangle\langle v_j| \otimes |v_r\rangle\langle v_s| \mapsto |v_j\rangle\langle v_i| \otimes |v_r\rangle\langle v_s|.
\end{align}
Similarly, we can define the partial transposition \(t_2\) for the second subsystem and extend it to a multi-system setting. The partial transposition has a natural connection with entanglement. For example, taking the operator $V_{(12)}$ and applying on it $t_2$, we get $dP^+_{12}$. The operator $P^+_{12}$ is just $|\psi^+_{12}\>\<\psi^+_{12}|$, where $|\psi^+_{12}\rangle=(1/\sqrt{d})\sum_{i,j=1}^d|i\rangle\otimes |i\rangle$ is the maximally entangled state in the computational basis of $\mathbb{C}^d \otimes \mathbb{C}^d$.

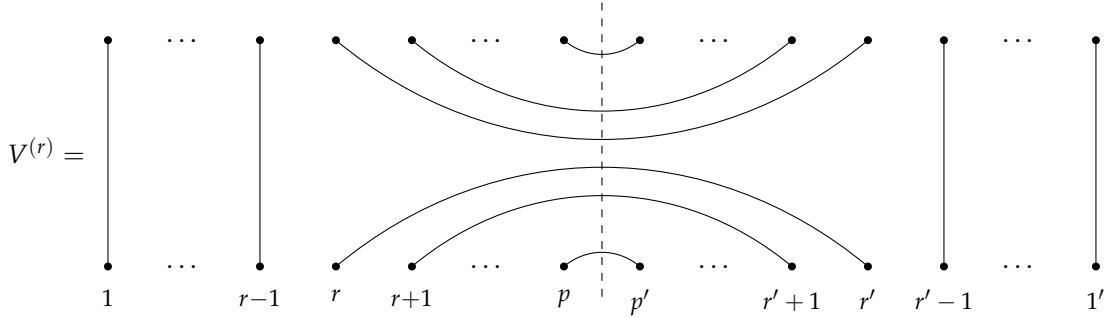
\begin{figure}[h!]
\centering
$V^{(r)} =$
\begin{tikzpicture}[scale=1, baseline=(current bounding box.center)]
  \def\yTop{3.0}
  \def\yBot{0}
  \def\dotRadius{1.5pt}
  \def\pSep{1.0}

  \foreach \x/\label in {
    0/1, 
    1/, 
    2/{r{-}1}, 
    3/r, 
    4/{r{+}1}, 
    5/, 
    6/{p}, 
    7/{p'}, 
    8/, 
    9/{r'+1}, 
    10/{r'}, 
    11/{r'-1},
    12/, 
    13/{1'}
  } {
    \ifx\label\empty
    \else
      \fill (\x*\pSep, \yBot) circle (\dotRadius);
    \fi
    \node[below=5pt] at (\x*\pSep, \yBot) {\small$\label$};
  }

  \node at (1*\pSep, \yBot) {$\dots$};
  \node at (5*\pSep, \yBot) {$\dots$};
  \node at (8*\pSep, \yBot) {$\dots$};
  \node at (12*\pSep, \yBot) {$\dots$};

  \foreach \x in {0,2,3,4,6,7,9,10,11,13} {
    \fill (\x*\pSep, \yTop) circle (\dotRadius);
  }
  \node at (1*\pSep, \yTop) {$\dots$};
  \node at (5*\pSep, \yTop) {$\dots$};
  \node at (8*\pSep, \yTop) {$\dots$};
  \node at (12*\pSep, \yTop) {$\dots$};

  \draw[dashed] (6.5*\pSep, \yTop + 0.5) -- (6.5*\pSep, \yBot - 0.5);

  \foreach \x in {0,2,11,13} {
    \draw (\x*\pSep, \yTop) -- (\x*\pSep, \yBot);
  }

  \draw[bend right=40] (3*\pSep,\yTop) to (10*\pSep,\yTop);
  \draw[bend right=40] (4*\pSep,\yTop) to (9*\pSep,\yTop);
  \draw[bend right=40] (6*\pSep,\yTop) to (7*\pSep,\yTop);

  \draw[bend left=40] (3*\pSep,\yBot) to (10*\pSep,\yBot);
  \draw[bend left=40] (4*\pSep,\yBot) to (9*\pSep,\yBot);
  \draw[bend left=40] (6*\pSep,\yBot) to (7*\pSep,\yBot);

\end{tikzpicture}
\caption{\small Diagram representation of the operator $V^{(r)}$ from equation~\eqref{eq:Vl}.  On the abstract level, this object is an element of the walled Brauer algebra $\mathcal{B}_{p,p}^\delta$ with $\delta=d$. On the representation space $(\mathbb{C}^d)^{\otimes 2p}$, the operator $V^{(r)}$ is an element of the algebra of the partially transposed permutation operators $\mathcal{A}_{p,p}^{d}$. }
\label{fig_Vr}
\end{figure}

Now, let us assume we are given $2p$ systems and we label them as $1,2,\ldots,p,p',p'-1,\ldots,1'$. As a result of applying partial transpositions $t_{p'},\ldots, t_{1'}$ to the natural representation of $\mathbb{C}[\mathfrak{S}_{2p}]$ we obtain the algebra of partially transposed operators serving as a matrix representation of the walled Brauer algebra $\mathcal{B}^d_{p,p}$:
\begin{align}
\label{eqn:a_p^k}
\mathcal{A}_{p,p}^{d} := \operatorname{span}_{\mathbb{C}}\{ V^{t_{p'}\circ t_{p'+1}\circ \cdots \circ t_{1'}}_{\pi} : \pi \in \mathfrak{S}_{2p} \},
\end{align}
where \(t_{p'}\circ t_{p'-1}\circ \cdots \circ t_{1'}\) represents the composition of the partial transpositions for the respective systems. Notice that the defined algebra is no longer a group algebra, since some elements are not invertible, for example, for $\mathcal{A}_{1,1}^d$, we have $V_{(1,1')}^{t_{1'}}V_{(1,1')}^{t_{1'}}=dV_{(1,1')}^{t_{1'}}$. 
In $\mathcal{A}_{p,p}^{d}$ we distinguish a special kind of objects, namely:
\begin{align}
\label{eq:Vl}
V^{(r)}&:=\id_{1,1'}\otimes\ldots\otimes \id_{r-1,r'-1}\otimes V_{(r,r')}^{t_{r'}}\otimes V_{(r+1,r'+1)}^{t_{r'+1}} \otimes \ldots  \otimes V_{(p,p')}^{t_{p'}}.
\end{align}
For the reader convenience the operator $V^{(r)}$ is visualized in Figure~\ref{fig_Vr}.  In particular, the operator $V^{(r)}$ from~\eqref{eq:Vl} can be viewed as a tensor product of $r$ projectors on maximally entangled states $P^+_{r,r'} \otimes P^+_{r+1,r'+1}\otimes \cdots \otimes  P^+_{p,p'}$, where we omitted identities for clarity.

The algebra $\mathcal{A}_{p,p}^{d}$ has a lot of interesting properties. For our further purposes, we recall here a generalised property from~\cite{studziński2025WBA}, namely the following contraction (or squeezing) lemma:
\begin{lemma}
\label{lemma1}
For an arbitrary operator $X\in \mathcal{A}_{p,p}^{d}$ and operator $V^{(r)}$, where $1\leq r\leq p$, given through equation~\eqref{eq:Vl} the following equality holds:
\begin{equation}
\label{eq:F1}
V^{(r)}XV^{(r)}=X_{p\setminus r, p\setminus r} V^{(r)},
\end{equation}
where $X_{p\setminus r, p\setminus r}\in \mathcal{A}^{d}_{p\setminus r, p\setminus r}$, and the symbol $p\setminus r$ means that the operators act non-trivially everywhere except the first $r$ arcs, counting from the wall.
\end{lemma}
The lemma above is intuitively clear, and follows from the composition rules within $\mathcal{A}^d_{p,p}$. However, finding an explicit form of the operator $X_{p\setminus r, p\setminus r}$, expressed in terms of matrix units of the subalgebra $\mathcal{A}^{d}_{p\setminus r, p\setminus r}$, is in general a very demanding task. We will discuss this problem in a more detailed way in the next sections.


Finally, we now formalize the notion of a group-adapted basis, which will play a
central role throughout the paper. We tailor the definition precisely to the notion of adaptedness appearing throughout the manuscript.

\begin{definition}
\label{def:group_adapted}
    Let $\mathcal{A}$ be an algebra carrying a natural action of
$\mathbb{C}[\mathfrak{S}_p] \times \mathbb{C}[\mathfrak{S}_p]$, induced by left and right
multiplication of permutation operators.
We say that a basis $\{B_\alpha\}$ of $\mathcal{A}$ is \emph{group-adapted}
(with respect to this action) if it is compatible with the decomposition
of $\mathcal{A}$ into irreducible subspaces under the joint
$\mathbb{C}[\mathfrak{S}_p] \times \mathbb{C}[\mathfrak{S}_p]$ action.
\end{definition}
More precisely, a basis is group-adapted if each basis element transforms
according to a fixed pair of irreducible representations
$(\mu,\nu)$ of $\mathfrak{S}_p$ under left and right multiplication, respectively,
and if the basis elements can be organized into blocks that span
invariant subspaces corresponding to these representation labels.
Equivalently, the basis resolves $\mathcal{A}$ into symmetry-compatible
components associated with irreducible matrix units of
$\mathbb{C}[\mathfrak{S}_p]$ from~\eqref{eqn:basis_Eij}.

In this sense, a group-adapted basis reflects the representation-theoretic
structure of the algebra and makes the underlying symmetry explicit at
the level of individual basis elements.

\section{Comparison with the known literature and our results}
\label{sec:comparison}
In this section, we summarise existing knowledge about constructing the group-adapted bases for the algebra of partially transposed permutation operators $\mathcal{A}_{p,p}^{d}$, and we briefly introduce the main idea for irreducible matrix basis construction for the whole algebra.
The  algebra $\mathcal{A}_{p,p}^{d}$ admits the following chain inclusion of two-sided ideals~\cite{Cox1,StudzinskiIEEE22}:
\begin{align}
\label{eqn:inclusions_of_m}
\mathcal{M}^{(p)} \subset \mathcal{M}^{(p-1)} \subset \cdots \subset \mathcal{M}^{(1)} \subset \mathcal{M}^{(0)} \equiv \mathcal{A}^{d}_{p,p},
\end{align}
where for $0\leq r\leq p$. For fixed $r$ the ideal  $\mathcal{M}^{(r)}$ is generated as
\begin{align}
\label{eqn:ideal_m}
\mathcal{M}^{(r)}=\langle\{ V_\sigma \otimes V_{\sigma'} V^{(r)} V_{\pi}^\dagger \otimes V_{\pi'}^\dagger \ | \ \sigma,\sigma',\pi,\pi'\in \mathfrak{S}_p\}\rangle,
\end{align}
where $V^{(r)}$ is given through~\eqref{eq:Vl}. The goal is to find irreducible matrix units which are group-adapted with respect to $\mathbb{C}[\mathfrak{S}_p] \times \mathbb{C}[\mathfrak{S}_p]$ producing the following decomposition of $\mathcal{A}_{p,p}^{d}$ into ideals:
\begin{align}
\label{eqn:inclusions_of_m2}
\mathcal{M}^{(p)} \oplus \widetilde{\mathcal{M}}^{(p-1)} \oplus \cdots \oplus \widetilde{\mathcal{M}}^{(1)} \oplus \widetilde{\mathcal{M}}^{(0)} \equiv \mathcal{A}^{d}_{p,p}.
\end{align}
For example, the ideal $\widetilde{\mathcal{M}}^{(p-1)}$ can be viewed as orthogonal complement to $\mathcal{M}^{(p)}$ contained only in $\mathcal{M}^{(p-1)}$.

Now, we are ready to explain the main approach to constructing the group-adapted bases for $\mathcal{A}^d_{p,p}$ producing the decomposition given in~\eqref{eqn:inclusions_of_m2}.  The considered algebra has been extensively studied in the context of port-based teleportation (PBT)~\cite{ishizaka_asymptotic_2008,studzinski2017port}, where the case $\mathcal{A}^d_{p,1}$ is used. Constructed irreducible matrix units, in this case, have the natural property of being group-adapted with respect to $\mathfrak{S}_p \times \mathfrak{S}_1$, since $ \mathfrak{S}_1$ is trivial. Further, by considering the multi-PBT, the algebra $\mathcal{A}^d_{p,p}$ came into play. However, due to the teleportation protocol property, the knowledge of only irreducible matrix units in the highest ideal $\mathcal{M}^{(p)}$ was required. The constructed basis in $\mathcal{M}^{(p)}$ has a required group-adapted nature, this time with respect to $\mathfrak{S}_p \times \mathfrak{S}_p$. Finally, in the recent paper~\cite{studziński2025WBA}, authors derived $\mathfrak{S}_p \times \mathfrak{S}_p$ group-adapted basis in the second-highest ideal $\widetilde{\mathcal{M}}^{(p-1)}$. Construction of the group-adapted basis in $\widetilde{\mathcal{M}}^{(p-1)}$ is more demanding. Since $\mathcal{M}^{(p)} \subset \mathcal{M}^{(p-1)}$, it can turn out that composing elements consisting only of $p-1$ arcs, we can produce elements with $p$ arcs. To be more precise, we can always produce arcs by composing elements, but we never decrease their number. In other words, we do not have the structure of the direct sum.

The main idea of construction relies on the fact that the ideals $\mathcal{M}^{(p)}, \mathcal{M}^{(p-1)}$ can be written as the following linear spans:
\begin{align}
&\mathcal{M}^{(p)}=\operatorname{span}_{\mathbb{C}}\left\{\Big(\overrightarrow{E}_L^p \otimes \overleftarrow{E}_R^p\Big) V^{(p)} \Big(\overrightarrow{E}_L^p \otimes \overleftarrow{E}_R^p\Big)\right\},\label{eq:linspan1}\\
&\mathcal{M}^{(p-1)}=\operatorname{span}_{\mathbb{C}}\left\{\Big(\overrightarrow{E}_L^p \otimes \overleftarrow{E}_R^p\Big) V^{(p)} \Big(\overrightarrow{E}_L^p \otimes \overleftarrow{E}_R^p\Big), \Big(\overrightarrow{E}_L^p \otimes \overleftarrow{E}_R^p\Big) V^{(p-1)} \Big(\overrightarrow{E}_L^p \otimes \overleftarrow{E}_R^p\Big)\right\}.\label{eq:linspan2}
\end{align}
The above spanning property is true since every permutation operator $V_\sigma,V_{\sigma'},V_\pi,V_{\pi'}\in \mathfrak{S}_p$ can be written in terms of irreducible matrix units of $\mathbb{C}[\mathfrak{S}_p]$, according to~\eqref{eqn:basis_Eij}. The number of operators in~\eqref{eq:linspan1} and~\eqref{eq:linspan2} is, of course, too large and can be reduced; however, it is not required to present the main idea here. Then the basis construction in $\mathcal{M}^{(p)}$ is simple, since we just must orthonormalize operators of the form $F_{p,p}=\Big(\overrightarrow{E}_L^p \otimes \id\Big) V^{(p)} \Big(\overrightarrow{E}_L^p \otimes \id\Big)$.  
Introducing notation $F_{p-1,p-1}= \Big(\overrightarrow{E}_L^p \otimes \overleftarrow{E}_R^p\Big) V^{(p-1)} \Big(\overrightarrow{E}_L^p \otimes \overleftarrow{E}_R^p\Big)$, it was shown in~\cite{studziński2025WBA} that the irreducible matrix units spanning the ideal $\widetilde{\mathcal{M}}^{(p-1)}$ are of the form:
\begin{align}
\label{eq:opH}
H(p-1,\alpha,\alpha')=dF_{p-1,p-1}-F_{p,p},\quad H(p-1,\alpha,\alpha')H(p-1,\beta,\beta')=db(\alpha',\beta)H(p-1,\alpha,\beta'),
\end{align}
where $\alpha, \alpha', \beta, \beta'$ are irreducible representations of $\mathfrak{S}_{p-1}$.  As we can see from the composition law, these operators are still not orthonormal, but they already have the required property of being orthogonal to $\mathcal{M}^{(p)}$. It means we have $F_{p,p}\cdot H(p-1,\alpha,\alpha')=H(p-1,\alpha,\alpha')\cdot F_{p,p}=0$. To transform the operators $H(p-1,\alpha,\alpha')$ to the orthonormal form, as explained in~\cite{studziński2025WBA}, we must diagonalize the matrix $(b(\alpha',\beta))$ and then renormalize operators from~\eqref{eq:opH}. Such diagonalization is always possible by a unitary matrix $U$, which can be derived for the fixed value of $p$.
Notice that we use here a very simplified notation which does not explain all technical difficulties and flavours. To have full notation treatment at this stage, we refer the reader to paper~\cite{studziński2025WBA}.

One of the main ingredients leading us to operators~\eqref{eq:opH} was an efficient method of calculating all possible overlaps between the operators $F_{p,p}$ and $F_{p-1,p-1}$, giving matrix elements of the respective Gram matrix.  This requires knowledge about the objects of the form $V^{(p-1)}\Big(\overrightarrow{E}_L^p \otimes \overleftarrow{E}_R^p\Big) V^{(p-1)}$, which is due to Lemma~\ref{lemma1} equal to $X_{1,1}\otimes V^{(p-1)}$. The operator $X_{1,1}$ belongs to the $\mathcal{A}^d_{1,1}=\operatorname{span}_\mathbb{C}\left\{\id, V^{(1)}\right\}$, so it can be written as
\begin{align}
X_{1,1}=a\id +b V^{(1)}=a\id +bdP^+,
\end{align}
where the coefficients $a,b \in \mathbb{R}$ are known and depend only on group-theoretic quantities such as dimensions and multiplicities of irreducible representations of $\mathfrak{S}_p$ and $\mathfrak{S}_{p-1}$, $P^+$ is a projector on maximally entangled state, and $d$ is the dimension of single particle space. It is very easy to see that the orthonormal basis in $\mathcal{A}^d_{1,1}$ is just $B_d(1,1)=\{\mathbf{1}-P^{+},P^{+}\}$. This roughly explains the main technical difficulty from the previous works and gives us a hint on how to construct bases in the remaining parts of the algebra. We want to continue the process of calculating the group-adapted basis for the remaining ideals. For definiteness, let us take ideal $\mathcal{M}^{(p-2)}$, this ideal is the following linear span
\begin{align}
\mathcal{M}^{(p-2)}=\operatorname{span}_{\mathbb{C}}\left\{F_{p,p},F_{p-1,p-1}, F_{p-2,p-2}\right\},
\end{align}
where 
\begin{align}
F_{p-2,p-2}=\Big(\overrightarrow{E}_L^p \otimes \overleftarrow{E}_R^p\Big) V^{(p-2)} \Big(\overrightarrow{E}_L^p \otimes \overleftarrow{E}_R^p\Big).
\end{align}
This time we need not only to know the basis for $\mathcal{A}^d_{1,1}$ but also for $\mathcal{A}^d_{2,2}$, since we deal with objects of the form $V^{(p-2)}\Big(\overrightarrow{E}_L^p \otimes \overleftarrow{E}_R^p\Big) V^{(p-2)}=X_{2,2}\otimes V^{(p-2)}$, where $X_{2,2}\in \mathcal{A}^d_{2,2}$. It means that to construct the group-adapted basis in $\widetilde{\mathcal{M}}^{(p-2)}$, one needs to have the full characterisation of $\mathcal{A}^d_{1,1}$ and $\mathcal{A}^d_{2,2}$ together. In general, having the ideal $\widetilde{\mathcal{M}}^{(p-r)}$, to construct group-adapted irreducible matrix units within it, we need to have orthogonal bases for the chain of the walled Brauer algebras - $\mathcal{A}^d_{1,1}, \mathcal{A}^d_{2,2},\ldots, \mathcal{A}^d_{r-1,r-1}, \mathcal{A}^d_{r,r}$. It means the orthogonal basis in $\mathcal{A}^d_{r,r}$ depends on the orthogonal basis elements from all previous steps. In fact, such a basis element of $\mathcal{A}^d_{r,r}$, where $1\leq r\leq p$ can be written in general as
\begin{align}
\label{eq:Brr}
B_{r,r}=\sum_{i=0}^r \sum_{L,R}a_{L,R}^{(i)}\overrightarrow{E}_L^{r}\otimes \overleftarrow{E}_R^{r} V^{(i)}\overrightarrow{E}_L^{r}\otimes \overleftarrow{E}_R^{r},
\end{align}
where $a^{(i)}_{L,R}$ are some coefficients that are established when the procedure of orthogonalization of every layer $\mathcal{A}^d_{i,i}$ is carried out. Notice that for the ideal $\widetilde{\mathcal{M}}^{(0)}$  of $\mathcal{A}^d_{r,r}$ the group-adapted basis has a very simple form~\footnote{In fact, for fixed index $0\leq i\leq r$,  we should distinguish between ideals $\mathcal{M}^{(i)}$ defined for algebras $\mathcal{A}^d_{r,r}$ for different values of $r$. Nevertheless, as long as it is clear from the context, we do not introduce an additional notation here.}. Namely, due to relation~\eqref{eqn:basis_Eij}, all basis operators are linear combinations of $\overrightarrow{E}_L^{k}\otimes \overleftarrow{E}_R^{k}$. 

In this way, the construction of $B_{r,r}$ from~\eqref{eq:Brr} is reduced to determining the
finite family of coefficients $a^{(i)}_{L,R}$ produced by the
orthogonalisation of each layer $\mathcal{A}^d_{i,i}$. Although finding these
coefficients remains challenging in general, for fixed and small $p$ the problem is computationally feasible.   It is also clear that the irreducible matrix units $B_{r,r}$ are group-adapted with respect to the action of $\mathbb{C}[\mathfrak{S}_p] \times \mathbb{C}[\mathfrak{S}_p]$ - as it is explained in Definition~\ref{def:group_adapted}. Indeed, the left (respectively right) action of any element
$\overrightarrow{E}_L^{r}\otimes \overleftarrow{E}_R^{r} \in 
\mathbb{C}[\mathfrak{S}_p] \times \mathbb{C}[\mathfrak{S}_p]$,
due to relations~\ref{eqn:basis_Eij}, maps each basis element to a linear
combination of elements transforming under the same pair of irreducible
representations. This confirms that the basis is group-adapted with respect
to the $\mathbb{C}[\mathfrak{S}_p]\times\mathbb{C}[\mathfrak{S}_p]$ action. 

We formalize the above idea of the basis construction in the ideals $\widetilde{\mathcal{M}}^{(p)},\widetilde{\mathcal{M}}^{(p-1)},\ldots, \widetilde{\mathcal{M}}^{(0)}$ of $\mathcal{A}^d_{p,p}$ in the next section.

\section{The algorithm}
\label{sec:algo}
Our algorithm, described below, in its main parts, is equivalent to the orthogonalization of a certain class of Gram matrices, and it can be summarized as follows:

\begin{algorithm}[H]
\caption{Orthonormal group-adapted irreducible basis construction for the algebra $\mathcal A^d_{p,p}$}
\label{alg:main}
\begin{algorithmic}[1]
\Require Integer $p \ge 1$, dimension parameter $d$
\Ensure Orthonormal group-adapted irreducible basis of the algebra $\mathcal A^d_{p,p}$

\State \textbf{Step 1: Base level $\mathcal A^d_{1,1}$}
\State Construct the orthonormal basis $\{\,\mathbbm{1}-P^{+},\; P^{+}\,\}$.

\For{$k = 2$ to $p$}
    \State \textbf{(Iteration level $k$): Construct an adapted basis of $\mathcal A^d_{k,k}$.}

    \State \textbf{Step 2: Ideal $\mathcal M^{(k)}$}
    \State Take the orthonormal basis of the ideal $\mathcal M^{(k)}$
           previously constructed in~\cite{StudzinskiIEEE22}.

    \State \textbf{Step 3: Quotient layer $\widetilde{\mathcal M}^{(k-1)}$}
    \State Orthogonalize all $(k-1)$-arc operators against:
    \State \hspace{1.2em}(a) the $k$-arc operators from $\mathcal M^{(k)}$, and
    \State \hspace{1.2em}(b) among themselves.
    \State Use the orthonormal basis of $\mathcal A^d_{1,1}$ when $k=2$, or of
           $\mathcal A^d_{k-1,k-1}$ from the previous iteration.

    \State \textbf{Step 4: Lower layers down to $\widetilde{\mathcal M}^{(0)}$}
    \For{$j = k-2$ down to $0$}
        \State Construct a basis in $\widetilde{\mathcal M}^{(j)}$ orthogonal to:
        \State \hspace{1.2em}$\mathcal M^{(k)},\; \widetilde{\mathcal M}^{(k-1)},\;
        \ldots,\; \widetilde{\mathcal M}^{(j+1)}$.
    \EndFor

    \State After completing all layers, merge the orthonormal bases to obtain
    \State the adapted basis of $\mathcal A^d_{k,k}$ corresponding to
           decomposition~(18).
\EndFor

\State \Return Orthonormal basis for $\mathcal A^d_{p,p}$.
\end{algorithmic}
\end{algorithm}

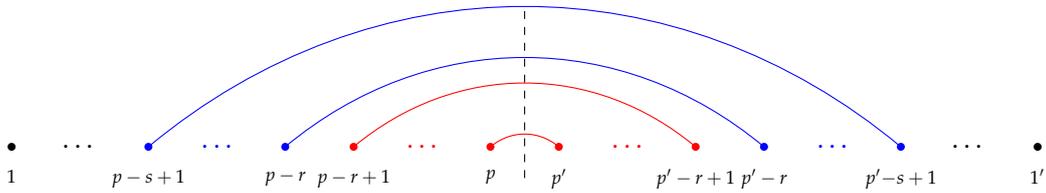
\begin{figure}[h!]
\centering
\begin{tikzpicture}[scale=1, baseline=(current bounding box.center)]
  \def\yTop{3.0}
  \def\yBot{0}
  \def\dotRadius{1.5pt}
  \def\pSep{0.9}

 \foreach \x/\label in {
    0/1, 
    1/, 
    2/{p-s+1}, 
    3/,
    4/{p-r}, 
    5/{p-r+1}, 
    6/, 
    7/{p}, 
    8/{p'}, 
    9/, 
    10/{p'-r+1},
    11/{p'-r}, 
    12/, 
    13/{p'{-}s+1},
    14/, 
    15/{1'}
} {
  \ifx\label\empty
  \else
    \ifnum\x=2
      \fill[blue] (\x*\pSep, \yBot) circle (\dotRadius);
    \else\ifnum\x=4
      \fill[blue] (\x*\pSep, \yBot) circle (\dotRadius);
    \else\ifnum\x=11
      \fill[blue] (\x*\pSep, \yBot) circle (\dotRadius);
    \else\ifnum\x=13
      \fill[blue] (\x*\pSep, \yBot) circle (\dotRadius);
    \else\ifnum\x=5
      \fill[red] (\x*\pSep, \yBot) circle (\dotRadius);
    \else\ifnum\x=7
      \fill[red] (\x*\pSep, \yBot) circle (\dotRadius);
    \else\ifnum\x=8
      \fill[red] (\x*\pSep, \yBot) circle (\dotRadius);
    \else\ifnum\x=10
      \fill[red] (\x*\pSep, \yBot) circle (\dotRadius);
    \else
      \fill (\x*\pSep, \yBot) circle (\dotRadius);
    \fi\fi\fi\fi\fi\fi\fi\fi
  \fi
  \node[below=5pt] at (\x*\pSep, \yBot) {\scriptsize$\label$};
}

  \node at (1*\pSep, \yBot) {$\dots$};
  \node at (3*\pSep, \yBot) {$\textcolor{blue}{\dots}$};
  \node at (6*\pSep, \yBot) {$\textcolor{red}{\dots}$};
  \node at (9*\pSep, \yBot) {$\textcolor{red}{\dots}$};
  \node at (12*\pSep, \yBot) {$\textcolor{blue}{\dots}$};
  \node at (14*\pSep, \yBot) {$\dots$};

  \draw[dashed] (7.5*\pSep, \yTop - 1.2) -- (7.5*\pSep, \yBot - 0.5);

  \draw[blue, bend left=40] (2*\pSep,\yBot) to (13*\pSep,\yBot);
  \draw[blue, bend left=40] (4*\pSep,\yBot) to (11*\pSep,\yBot);
  \draw[red, bend left=40] (5*\pSep,\yBot) to (10*\pSep,\yBot);
  \draw[red, bend left=40] (7*\pSep,\yBot) to (8*\pSep,\yBot);

\end{tikzpicture}
\caption{\small Graphic presents relation between non-trivial action of $V^{(r)},V^{(s)}$, and $V^{(s\setminus r)}$. The red arcs represent the domain for the operator $V^{(r)}$. The red arcs, together with the blue arcs, give us a domain for $V^{(s)}$. From this, it is easy to deduce that the blue arcs are for $V^{(s\setminus r)}$. For the clarity of the figure, we draw here only the bottom part of the considered elements.}
\label{fig_Vrsets}
\end{figure}

We observe that each subsequent step in algorithm~\ref{alg:main} utilizes the results from all preceding steps. The mentioned procedure of orthogonalization requires computing matrix elements of the respective Gram matrices. These matrix elements are equal to the traces computed between objects spanning the respective ideals $\mathcal{M}^{(n)}, \mathcal{M}^{(n-1)},\ldots,\mathcal{M}^{(0)}$, where $1\leq m \leq p$. As we saw, to produce the basis in the final layer, i.e., when $m=p$, one must use the basis produced in all previous steps. 
To understand all steps of the algorithm from the technical point of view, without loss of generality, let us assume that our goal is to orthogonalize objects consisting of $s$ arcs with objects consisting of $r$ arcs, where $r<s$ in the $p:p$ case. The idea of orthonormalization requires calculating the following overlaps:
\begin{align}
\label{eq:1}
\tr\left[\left(\overrightarrow{E}_L^p \otimes \overleftarrow{E}_R^p V^{(r)}\overrightarrow{E}_L^p \otimes \overleftarrow{E}_R^p\right)\left(\overrightarrow{E}_L^p \otimes \overleftarrow{E}_R^p V^{(s)}\overrightarrow{E}_L^p \otimes \overleftarrow{E}_R^p\right)\right]=\tr\left[\left(\overrightarrow{E}_L^p \otimes \overleftarrow{E}_R^p V^{(r)}\overrightarrow{E}_L^p \otimes \overleftarrow{E}_R^p  V^{(s)}\right)\right],
\end{align}
where we used the composition law~\eqref{eqn:basis_Eij} for the irreducible matrix units in $\mathbb{C}[\mathfrak{S}_p]$. Notice that due to the mentioned composition law~\eqref{eqn:basis_Eij}, many of the matrix entries are equal to 0. This greatly simplifies computations throughout the algorithm, as numerous trace terms are provided implicitly and do not require explicit evaluation. In the next step, let us observe that the operator $V^{(s)}$ can be written as $V^{(r)}V^{(s\setminus r)}$, where the operator $V^{(s\setminus r)}$ represents the remaining arcs, see Figure~\ref{fig_Vrsets}.

In particular, we have $[V^{(r)},V^{(s\setminus r)}]=0$. We can continue calculating~\eqref{eq:1} as
\begin{align}
\label{eq:cacl}
\tr\left[\left(\overrightarrow{E}_L^p \otimes \overleftarrow{E}_R^p V^{(r)}\overrightarrow{E}_L^p \otimes \overleftarrow{E}_R^p  V^{(s)}\right)\right]=\tr\left[\left(\overrightarrow{E}_L^p \otimes \overleftarrow{E}_R^p V^{(r)}\overrightarrow{E}_L^p \otimes \overleftarrow{E}_R^p  V^{(r)}\right)V^{(s\setminus r)}\right].
\end{align}
For the element $V^{(r)}E_L^p \otimes E_R^p  V^{(r)}$ we apply Lemma~\ref{lemma1}, and  write it in the following form
\begin{align}
 V^{(r)}\overrightarrow{E}_L^p \otimes \overleftarrow{E}_R^p  V^{(r)}=X_{p\setminus r,p\setminus r}\otimes V^{(r)}, \quad \text{where} \quad    X_{p\setminus r,p\setminus r}\in \mathcal{A}^d_{p-r,p-r},
\end{align}
Assuming we know the orthonormal irreducible matrix units $F_{p\setminus r, p\setminus r}$ for the algebra $\mathcal{A}^d_{p-r,p-r}$ from the previous steps of the algorithm, which is given in general form as
\begin{align}
\label{eq:basis}
F_{p\setminus r, p\setminus r}&=\sum_{l=0}^{p-r}\sum_{L,R}a^{(l)}_{L,R}\overrightarrow{E}_L^{p\setminus r}\otimes \overleftarrow{E}_R^{p\setminus r} V^{(l)}\overrightarrow{E}_L^{p\setminus r}\otimes \overleftarrow{E}_R^{p\setminus r}\\
&\equiv \overrightarrow{E}_L^{p\setminus r}\otimes \overleftarrow{E}_R^{p\setminus r} V^{(l)}\overrightarrow{E}_L^{p\setminus r}\otimes \overleftarrow{E}_R^{p\setminus r},
\end{align}
where $a^{(l)}_{L,R}$ are some coefficients that are established when the procedure of orthogonalization of $\mathcal{A}^d_{p-r,p-r}$ is carried out. Notice that the operator $V^{(l)}$ acts on systems $(1,\ldots,p-r)\times (p+r+1,\ldots, 2p)$ when we embed algebra $\mathcal{A}^d_{p-r,p-r}$ into $\mathcal{A}^d_{p,p}$. For simplicity of the notation, we do not introduce here a new symbol for $V^{(l)}$.  Additionally, we used a simplified notation for summations.  The lower indices $L,R$ include summation over all indices included in the operators $ \overrightarrow{E}_L^{p\setminus r}$ and $ E_R^{p\setminus r}$ weighted by some coefficients appearing in the linear combination. This approach is fully justified here, since our goal is to present the principle of operation of the algorithm, and its application will be illustrated in detail in the next section. Keeping this in mind, after expressing the operator $X_{p\setminus r,p\setminus r}$ in the basis~\eqref{eq:basis}, we continue calculating expression~\eqref{eq:cacl}:
\begin{align}
\label{line0}
\tr\left[\overrightarrow{E}_L^p \otimes \overleftarrow{E}_R^p V^{(r)}F_{p\setminus r, p\setminus r}V^{(s\setminus r)}\right]&=\tr\left(V^{(r)}\overrightarrow{E}_L^p \otimes \overleftarrow{E}_R^p V^{(r)}F_{p\setminus r, p\setminus r}V^{(s\setminus r)}\right)\\ \label{line1}
&=\tr\left[V^{(r)}(\overrightarrow{E}_L^p \otimes \overleftarrow{E}_R^p) V^{(r)}(\overrightarrow{E}_L^{p\setminus r}\otimes \overleftarrow{E}_R^{p\setminus r}) V^{(l)}(\overrightarrow{E}_L^{p\setminus r}\otimes \overleftarrow{E}_R^{p\setminus r})V^{(s\setminus r)}\right]\\ \label{line2a}
&=\tr\left[(\overrightarrow{E}_L^{p\setminus r}\otimes \overleftarrow{E}_R^{p\setminus r})(\overrightarrow{E}_L^p \otimes \overleftarrow{E}_R^p) (\overrightarrow{E}_L^{p\setminus r}\otimes \overleftarrow{E}_R^{p\setminus r}) V^{(l)}V^{(s\setminus r)}V^{(r)}\right]\\ \label{line3b}
&=\tr\left[(\overrightarrow{E}_L^p \otimes \overleftarrow{E}_R^p) V^{(s+l')}\right].
\end{align}
In line~\eqref{line0} we use property $V^{(r)}V^{(r)}=d^r V^{(r)}$, where we omit the constant factor for transparency, and commutation relations $[V^{(r)}, V^{(s\setminus r)}]=0$, $[V^{(r)},F_{p\setminus r, p\setminus r}]=0$.
In the equation~\eqref{line1} we substitute the explicit form of the irreducible matrix units $F_{p\setminus r, p\setminus r}$ from~\eqref{eq:basis}. To obtain expression from~\eqref{line2a}, we observe that $(\overrightarrow{E}_L^{p\setminus r}\otimes \overleftarrow{E}_R^{p\setminus r}) V^{(l)}(\overrightarrow{E}_L^{p\setminus r}\otimes \overleftarrow{E}_R^{p\setminus r})$ acts non-trivially only on the part $p\setminus r$, so it commutes with $V^{(r)}$. In the next step, however, we use the cyclicity of the full trace, property $V^{(r)}V^{(r)}=d^r V^{(r)}$, where we omit the constant factor for transparency, and $V^{(r)}V^{(l)}=V^{(r+l)}$. Finally, in line~\eqref{line3b}, since $\overrightarrow{E}^p_L, \overleftarrow{E}^p_R\in \mathbb{C}[\mathfrak{S}_p]$ and $\overrightarrow{E}_L^{p\setminus r},\overleftarrow{E}_R^{p\setminus r}\in \mathbb{C}[\mathfrak{S}_{p-r}]$, we use the absorption property $\overrightarrow{E}_L^p\overrightarrow{E}_L^{p\setminus r} \in \mathbb{C}[\mathfrak{S}_p]$, and $\overleftarrow{E}_R^p\overleftarrow{E}_R^{p\setminus r} \in \mathbb{C}[\mathfrak{S}_p]$.  Next, to get expression~\eqref{line3b}, we observe that $V^{(l)}V^{s\setminus r}V^{(r)}=V^{(s+l')}$, where $l'$ is an index indicating those $l$ for which the operator $V^{(l)}$ acts non-trivially outside the first $s$ registers counting from the wall in the both directions. Otherwise, we apply multiplication rule $V^{(l)}V^{(s)}=d^{l}V^{(s)}$.  Finally, applying the trace cyclicity, we write
\begin{align}
\tr\left[\overrightarrow{E}_L^p \otimes \overleftarrow{E}_R^p V^{(r)}F_{p\setminus r, p\setminus r}V^{(s\setminus r)}\right]&=\tr\left[(\overrightarrow{E}_L^p \otimes \overleftarrow{E}_R^p) V^{(s+l')}\right]\\
&=\tr\left[V^{(s+l')}(\overrightarrow{E}_L^p \otimes \overleftarrow{E}_R^p) V^{(s+l')}\right].
\end{align}
Let us observe that we can go with the partial trace in the bracket and write
\begin{align}
\label{partial_traces}
\tr\left[\overrightarrow{E}_L^p \otimes \overleftarrow{E}_R^p V^{(r)}F_{p\setminus r, p\setminus r}V^{(s\setminus r)}\right]=\tr\left[\left(\tr_{p\setminus(s+l')}(\overrightarrow{E}_L^p)\otimes \tr_{p\setminus(s+l')}(\overleftarrow{E}_R^p)\right)V^{(s+l')} \right].
\end{align}
From expression~\eqref{partial_traces}, we see that there is a problem with calculating both partial traces $\tr_{p\setminus(s+l')}(\overrightarrow{E}_L^p)$ and $\tr_{p\setminus(s+l')}(\overleftarrow{E}_R^p)$. Namely, for fixed $l'$ and $s$, the first partial trace is taken over the first subsystems, which is not compatible with the Young-Yamanouchi basis construction described briefly in Section~\ref{sec:tech_intro}. For the second partial trace, the situation is analogous. In other words, a closed-form expression for the partial trace can only be derived when it is computed iteratively over the last subsystems, starting from the final one. To overcome the problem, instead of dealing with the operators $\overrightarrow{E}_L^p,\overleftarrow{E}_R^p$, we need to consider 
\begin{align}
\label{sandwich}
\overrightarrow{E}_L^p=V_\sigma \overleftarrow{E}_L^p V_{\sigma^{-1}},\qquad \overleftarrow{E}_R^p=V_\sigma \overrightarrow{E}_R^p V_{\sigma^{-1}}
\end{align}
where $\sigma \in \mathfrak{S}_p$ permutes systems in the following order $1 \leftrightarrow p, 2 \leftrightarrow (p-1)$ and so on. Notice that the sandwiching in~\eqref{sandwich} can be expanded by applying formula~\eqref{eq:actionVonE}. It means we can write the right-hand side of~\eqref{partial_traces} as a linear combination of terms consisting of partial traces, which we can easily deal with. Let us apply~\eqref{sandwich} to~\eqref{partial_traces}, but for transparency, omit all coefficients appearing in front of the full trace as a result:
\begin{align}
\label{partial_traces2}
\tr\left[\overrightarrow{E}_L^p \otimes \overleftarrow{E}_R^p V^{(r)}F_{p\setminus r, p\setminus r}V^{(s\setminus r)}\right]&=\tr\left[\left(\tr_{p\setminus(s+l')}(\overleftarrow{E}_L^p)\otimes \tr_{p\setminus(s+l')}(\overrightarrow{E}_R^p)\right)V^{(s+l')} \right]\\ \label{line2}
&=\tr\left[\left(\tr_{p\setminus(s+l')}(\overleftarrow{E}_L^p)\tr_{p\setminus(s+l')}(\overrightarrow{E}_R^p)^T\otimes \id\right)V^{(s+l')}\right]\\ 
&=\tr\left[\tr_{p\setminus(s+l')}(\overleftarrow{E}_L^p)\tr_{p\setminus(s+l')}(\overrightarrow{E}_R^p)^T\right]. \label{line3}
\end{align}
After evaluating the partial traces, the results act non-trivially on the same systems as the operator $V^{(s+l')}$. Due to this property, we can apply the 'ping-pong' trick $(\id \otimes X)|\psi^+\>=(X^T\otimes \id)|\psi^+\>$, and trace out all systems on the right-hand side of the wall. It means that in the whole algorithm, the crucial role is played by the numbers defined in the line~\eqref{line3}. 

Although Algorithm~\ref{alg:main} is fully constructive, its explicit
closed-form outputs become rapidly complex as the layer index $i$ grows.
The main computational bottlenecks are:
\begin{enumerate}[(1)]
    \item assembling the Gram matrices from Hilbert--Schmidt overlaps of the
(overcomplete) generators in each layer $\mathcal{A}^{i,i}_d$ (see Section~\ref{sec:decomposition} for more details),
\item performing a rank-revealing orthogonalisation to extract an independent,
group-adapted basis (see Section~\ref{sec:decomposition} for more details), and
\item expanding contraction outputs $X_{p\setminus r,\,p\setminus r}$
in a chosen spanning family of $\mathcal{A}^{p-r,p-r}_d$.
\end{enumerate}

The group-adapted structure yields strong
patterns---in particular, block decompositions of Gram matrices according
to irreducible labels and many vanishing entries due to symmetry/selection
rules---the coefficients produced by orthogonalisation involve inverses and
eigenstructure of these blocks, which typically leads to expressions that
are too lengthy to present in closed form beyond small $i$.

For this reason, we provide fully explicit formulae for the first layers
(e.g.\ $\mathcal{A}^{2,2}_d$ and $\mathcal{A}^{3,3}_d$ and the corresponding
$\widetilde{\mathcal{M}}^{(i)}$ ideals) as representative examples, while
for general $i$ we view the construction primarily as an effective
finite-dimensional linear-algebra procedure at fixed system size $p$.

\section{Decomposition of the algebra $\mathcal{A}^d_{p,p}$ into direct sum of $\widetilde{\mathcal{M}}^{(p)}$ and $\widetilde{\mathcal{M}}^{(p-1)}$}
\label{sec:decomposition}

The operators $V^{(r)}:r=0,1,\ldots,p$ given in~\eqref{eq:Vl}, which span every ideal $\mathcal{M}^{(r)}$ in~\eqref{eqn:ideal_m} are essential projectors. It means they satisfy the relation $
V^{(r)}V^{(r)}=d^{r}V^{(r)}$ but they are  not
orthogonal therefore they induce only decomposition of $\mathcal{A}^d_{p,p}$ into chain of inclusions of ideals~\eqref{eqn:inclusions_of_m}. To get rid of this problem, we introduce a new set of projectors, which are orthogonal by construction and yield a new decomposition of $\mathcal{A}^d_{p,p}$ into the direct sum of ideals, as presented in~\eqref{eqn:inclusions_of_m2}. The newly proposed set of projectors yields the two highest ideals in~\eqref{eqn:inclusions_of_m2} in an efficient and elegant manner. Importantly, the underlying construction is not restricted to this case and can, in principle, be extended to the lowest ideals. Such an extension requires additional technical developments, which are beyond the scope of the present paper and will be explored in future work.

\begin{definition}
\label{def:Q}
For $0,1,,\ldots,p$ we define the following set of operators:
\begin{align}
\label{eq:Qs}
&Q^{(p)}=\frac{1}{d^{p}}V^{(p)},\quad Q^{(p-1)}=\frac{1}{d^{p-1}}V^{(p-1)}-\frac{1}{d^{p}}V^{(p)},\ldots \\ 
&Q^{(k)}=\frac{1}{d^{k}}V^{(k)}-\frac{1}{d^{k+1}}V^{(k+1)},\ldots, Q^{(1)}=
\frac{1}{d}V^{(1)}-\frac{1}{d^{2}}V^{(2)},\quad Q^{(0)}=\mathbf{1}-\frac{1}{d}
V^{(1)},
\end{align}
where the operators $V^{(r)}:r=0,1,\ldots,p$ are given in~\eqref{eq:Vl}.
\end{definition}

Directly the multiplication rules for $V^{(r)}:r=0,1,\ldots,p$ we can derive multiplication rules for all operators $Q^{(k)}:k=0,1,\ldots,p$.

\begin{proposition}
The operators $Q^{(k)}:k=0,1,\ldots,p$ given in Definition~\ref{def:Q}, satisfy the following multiplication and summation rules:
\begin{equation}
Q^{(k)}Q^{(l)}=\delta ^{kl}Q^{(k)},\quad \sum_{k=0}^{p}Q^{(k)}=\mathbf{1}_{(
\mathbb{C}^{d})^{\otimes 2p}},
\end{equation}
The operators $Q^{(k)}:k=0,1,\ldots,p$ form a complete set of orthogonal projectors
on the space $(\mathbb{C}^{d})^{\otimes 2p}$, therefore we have
\begin{equation}
(\mathbb{C}^{d})^{\otimes 2p}=\bigoplus _{k=0}^{p}\operatorname{supp}(Q^{(k)}),
\end{equation}
where 
\begin{equation}
\dim \left( \operatorname{supp}(Q^{(k)})\right)
=\operatorname{Tr}(Q^{(k)})=d^{2(p-k-1)}(d^{2}-1):k=0,1,\ldots,p-1,\quad \operatorname{Tr}(Q^{(p)})=1.
\end{equation}
\end{proposition}
The statement of the above proposition can be verified by straightforward calculations, using Definition~\ref{def:Q} and the properties of orthogonal projectors.

It is clear that the transformation between operators $Q^{(k)}$ and $V^{(k)}$ is invertible. Namely, we have
\begin{equation}
\forall l=1,2,\ldots,p \quad d^l\left(Q^{(l)}+Q^{(l+1)}+\cdots +Q^{(p-1)}+Q^{(p)}\right)=V^{(l)}
\end{equation}
and
\begin{equation}
\sum_{k=1}^pQ^{(k)}=\mathbf{1}_{(\mathbb{C}^d)^{\otimes 2p}}=V^{(0)}.
\end{equation}
It is also clear that $p+1$ projectors $Q^{(k)}:k=0,1,\ldots,p$ are linearly independent (because they are orthogonal), and they are linearly generated by $p+1$ operators $V^{(k)}:k=0,1,\ldots,p$. Therefore, $p+1$ operators $V^{(k)}$ are also linearly independent because a set of $p+1$ linearly dependent operators cannot linearly generate another set of $p+1$ linearly independent operators. Therefore, we have:

\begin{corollary}
\label{cor:1}
The algebra $\mathcal{A}_{p,p}^d$ is spanned by the operators $Q^{(k)}:k=0,1,\ldots,p$ given in Definition~\ref{def:Q}:
\begin{equation}
\begin{split}
\mathcal{A}_{p,p}^d=\operatorname{span}_{\mathbb{C}}\{&\overrightarrow{E}_{ij}^\mu\otimes \overleftarrow{E}^\nu_{kl} Q^{(p)}\overrightarrow{E}_{ab}^\xi\otimes \overleftarrow{E}^\eta_{cd},\quad \overrightarrow{E}_{ij}^\mu\otimes \overleftarrow{E}^\nu_{kl} Q^{(p-1)}\overrightarrow{E}_{ab}^\xi\otimes \overleftarrow{E}^\eta_{cd},\ldots,\\
&\overrightarrow{E}_{ij}^\mu\otimes \overleftarrow{E}^\nu_{kl} Q^{(1)}\overrightarrow{E}_{ab}^\xi\otimes \overleftarrow{E}^\eta_{cd},\quad \overrightarrow{E}_{ij}^\mu\otimes \overleftarrow{E}^\nu_{kl} Q^{(0)}\overrightarrow{E}_{ab}^\xi\otimes \overleftarrow{E}^\eta_{cd}\},
\end{split}
\end{equation}
where $\overrightarrow{E}_{ij}^\mu,\overleftarrow{E}^\nu_{kl},\overrightarrow{E}_{ab}^\xi,\overleftarrow{E}^\eta_{cd}$ are irreducible matrix units of $\mathbb{C}[\mathfrak{S}_p]$  from~\eqref{eqn:basis_Eij} constructed in a proper order indicated by the arrows.
\end{corollary}

For further simplification in this section, we omit the arrows above the operators $E_{ij}^\mu$, remembering that they are constructed in opposite directions on both sides of the wall.
Also, from now on, we will be using the spanning set for the algebra $\mathcal{A}_{p,p}^d$ from Corollary~\ref{cor:1}.
Using notation from equations~\eqref{eq:notationPRIR1} and~\eqref{eq:notationPRIR2} let us multiply both sides of~\eqref{eq:WBA_praca} from Theorem~\ref{thmWBA} in Appendix~\ref{app:A} by
\begin{equation}
\frac{1}{d^{p-1}}\left(\mathbf{1}-\frac{1}{d}V^{t_{1'}}_{(1,1')}\right).
\end{equation}%
Then  taking into account that
\begin{equation}
Q^{(p-1)}=\frac{1}{d^{p-1}}V^{(p-1)}\left(\mathbf{1}-\frac{1}{d}V^{t_{1'}}_{(1,1')}\right) ,\quad Q^{(p)}\left( \mathbf{1}-\frac{1}{d}V^{t_{1'}}_{(1,1')}\right) =0,
\end{equation}
we formulate the variant of  Lemma~\ref{lemWBA} from Appendix~\ref{app:A}. Namely, we have the following

\begin{proposition}
Orthogonal projector $Q^{(p-1)}$, given through Definition~\ref{def:Q}, satisfies relation 
\begin{equation}
Q^{(p-1)}E_{I_{\alpha }J_{\alpha ^{\prime }}}^{\mu }\otimes E_{K_{\beta
}L_{\beta ^{\prime }}}^{\nu }Q^{(p-1)}=\frac{1}{d^{p}(d^{2}-1)}\delta
_{I_{\alpha }K_{\beta }}\delta _{J_{\alpha ^{\prime }}L_{\beta ^{\prime
}}}\left( d\frac{m_{\mu }m_{\nu }\delta ^{\alpha \alpha ^{\prime }}}{
m_{\alpha }}-m_{\nu }\delta ^{\mu \nu }\right) Q^{(p-1)},
\end{equation}
where $E_{I_{\alpha }J_{\alpha ^{\prime }}}^{\mu }, E_{K_{\beta
}L_{\beta ^{\prime }}}^{\nu }$ are irreducible matrix units for $\mathbb{C}[\mathfrak{S}_p]$ from~\eqref{eqn:basis_Eij} written in the notation from~\eqref{eq:notationPRIR1},~\eqref{eq:notationPRIR2}.
\end{proposition}

From this, using similar methods, we deduce

\begin{corollary}
\label{Cor:6}
The subspaces
\begin{align}
&\widetilde{\mathcal{M}}^{(p)}=\operatorname{span}_{\mathbb{C}}\{\mathbf{1}\otimes E_{cd}^{\eta }Q^{(p)}E_{ij}^{\mu }\otimes \mathbf{1}:\quad \eta ,\mu \in \widehat{\mathfrak{S}}_p\},\\
&\widetilde{\mathcal{M}}^{(p-1)}=\operatorname{span}_{\mathbb{C}}\{E_{ab}^{\xi }\otimes E_{cd}^{\eta }Q^{(p-1)}E_{ij}^{\mu }\otimes
E_{kl}^{\nu }:\quad \xi ,\eta ,\mu ,\nu \in \widehat{\mathfrak{S}}_p\},\label{eq:tildeMp1}
\end{align}
where $\widehat{\mathfrak{S}}_p$ is the set of all irreps of $\mathfrak{S}_p$,
are subalgeras of $\mathcal{A}^d_{p,p}$, such that 
\begin{equation}
\widetilde{\mathcal{M}}^{(p)}\widetilde{\mathcal{M}}^{(p-1)}=0.
\end{equation}
This implies that they can be viewed as two orthogonal ideals contained in $\mathcal{A}^d_{p,p}$:
\begin{equation}
\widetilde{\mathcal{M}}^{(p)}\cup \widetilde{\mathcal{M}}^{(p-1)}=\widetilde{\mathcal{M}}^{(p)}\oplus \widetilde{\mathcal{M}}^{(p-1)}.
\end{equation}
\end{corollary}
The ideal $\widetilde{\mathcal{M}}^{(p-1)}$, although defined via the projector $Q^{(p-1)}$ instead of the essential projector $V^{(p-1)}$, may be described more precisely in a manner similar to Proposition 11 of~\cite{studziński2025WBA}. Namely, the next proposition specifies the constraints under which the spanning operators for the ideal $\widetilde{\mathcal{M}}^{(p-1)}$ from Corollary~\ref{cor:1} are non-zero. 

\begin{proposition}
\label{Prop:7}
An element $E_{ab}^{\xi }\otimes E_{cd}^{\eta }Q^{(p-1)}E_{ij}^{\mu }\otimes
E_{kl}^{\nu }\in \widetilde{\mathcal{M}}^{(p-1)}$ is non-zero if and only if it is of the form in the notation~\eqref{eq:notationPRIR1}:
\begin{equation}
\label{eq:operators}
E_{I_{\gamma }A_{\kappa }}^{\xi }\otimes E_{K_{\eta }A_{\kappa }}^{\eta
}Q^{(p-1)}E_{B_{\tau }J_{\alpha }}^{\mu }\otimes E_{B_{\tau }L_{\beta }}^{\nu
},
\end{equation}
where 
\begin{equation}
\xi \sim _{\square }\eta
 \quad \wedge \quad  \mu \sim _{\square }\nu
\end{equation}
and
\begin{equation}
A_{\kappa }=\left( 
\begin{array}{c}
\kappa \\ 
a_{\kappa }
\end{array}
\right) ,\quad B_{\tau }=\left( 
\begin{array}{c}
\tau \\ 
b_{\tau }
\end{array}
\right) :\quad \kappa = \xi-\Box \wedge \kappa = \eta-\Box \quad \wedge
\quad \tau = \mu-\Box \wedge \tau = \nu-\Box.
\end{equation}
It means operators~\eqref{eq:operators}, for fixed $\xi,\eta,\mu,\nu \vdash p$ depend on unique indices $\kappa,\tau \vdash (p-1)$ and
do not depend on the remaining indices $a_{\kappa }$ and $b_{\tau }$.
\end{proposition}
The operators from  Proposition~\ref{Prop:7} play an important role in the remaining part of the section, so we introduce the following simplifying notation:
\begin{definition}
\label{Not:1}
For the operators from Proposition~\ref{Prop:7}, we introduce the following notation:
\begin{equation}
\widehat{G}_{\Gamma \Delta }^{(p-1)}\equiv\widehat{G}_{\tiny{\left[ 
\begin{array}{cc}
\xi  & \eta  \\ 
I_{\gamma } & K_{\eta }
\end{array}
,\kappa \right] ,\left[ 
\begin{array}{cc}
\mu  & \nu  \\ 
J_{\alpha } & L_{\beta }
\end{array}
,\tau \right] }}^{(p-1)}:=E_{I_{\gamma }A_{\kappa }}^{\xi }\otimes E_{K_{\eta
}A_{\kappa }}^{\eta }Q^{p-1}E_{B_{\tau }J_{\alpha }}^{\mu }\otimes
E_{B_{\tau }L_{\beta }}^{\nu }.
\end{equation}
Each multi-index 
\begin{equation}
\Gamma =\left[ 
\begin{array}{cc}
\xi  & \eta  \\ 
I_{\gamma } & K_{\eta }
\end{array}
,\kappa \right] ,\quad \Delta =\left[ 
\begin{array}{cc}
\mu  & \nu  \\ 
J_{\alpha } & L_{\beta }
\end{array}
,\tau \right] 
\end{equation}
in $\widehat{G}_{\Gamma \Delta }^{(p-1)}$ contains five sub-indices.
\end{definition}
Having Definition~\ref{Not:1}, we can reasonably easily formulate the multiplication rule for the operators~\eqref{eq:operators} from Proposition~\ref{Prop:7}.
\begin{proposition}
\label{prop:multiB}
For operators the $\widehat{G}_{\Gamma \Delta }^{(p-1)}, \widehat{G}_{\Lambda \Pi }^{(p-1)}$ given through Definition~\ref{Not:1}, with multi-indices
\begin{equation}
\Gamma =\left[ 
\begin{array}{cc}
\xi  & \eta  \\ 
I_{\gamma } & K_{\eta }
\end{array}
,\kappa \right] ,\quad \Delta =\left[ 
\begin{array}{cc}
\mu  & \nu  \\ 
J_{\alpha } & L_{\beta }
\end{array}
,\tau \right],\quad  \Lambda=\left[ 
\begin{array}{cc}
\mu ^{\prime } & \nu ^{\prime } \\ 
I_{\alpha ^{\prime }} & L_{\beta ^{\prime }}
\end{array}
,\gamma \right],\quad 
\Pi=\left[ 
\begin{array}{cc}
\kappa  & \varsigma  \\ 
S & T
\end{array}
,\tau \right],
\end{equation}
we have the following composition rule 
\begin{equation}
\label{eq:compB}
\widehat{G}_{\Gamma \Delta }^{(p-1)}\widehat{G}_{\Lambda \Pi }^{(p-1)}=\widehat{B
}_{\Delta \Lambda }^{(p-1)}\widehat{G}_{\Gamma \Pi }^{(p-1)}.
\end{equation}
The Gram matrix $\widehat{B}^{(p-1)}$ defines the above multiplication in the ideal 
$\widetilde{\mathcal{M}}^{(p-1)}$ and has the following entries
\begin{equation}
\label{eq:matB}
\widehat{B}_{\Delta \Lambda }^{(p-1)}=\widehat{B}_{\tiny{\left[ 
\begin{array}{cc}
\mu  & \nu  \\ 
J_{\alpha } & K_{\beta }
\end{array}
,\omega \right] ,\left[ 
\begin{array}{cc}
\mu ^{\prime } & \nu ^{\prime } \\ 
I_{\alpha ^{\prime }} & L_{\beta ^{\prime }}
\end{array}
,\gamma \right] }}^{(p-1)}
=\frac{1}{d^{p}(d^{2}-1)}\delta ^{\mu \mu ^{\prime }}\delta ^{\nu \nu
^{\prime }}\delta _{J_{\alpha }I_{\alpha ^{\prime }}}\delta _{K_{\beta
}L_{\beta ^{\prime }}}\left[ d\frac{m_{\mu }m_{\nu }}{m_{\omega }}\delta
^{\omega \gamma }-m_{\mu }\delta ^{\mu \nu }\right].
\end{equation}
\end{proposition}
The matrix $\widehat{B}^{(p-1)}$ from equation~\eqref{eq:matB} has several interesting properties that define further properties of the multiplication from Proposition~\ref{prop:multiB} in the ideal $\widetilde{\mathcal{M}}^{(p-1)}$.
\begin{remark}
The matrix multiplication from~\eqref{eq:compB} in the ideal $\widetilde{\mathcal{M}}^{(p-1)}$, we call quasi-matrix
multiplication. It becomes pure matrix multiplication if 
\begin{equation}
\widehat{B}_{\Delta \Lambda }^{(p-1)}=\delta _{\Delta \Lambda }.
\end{equation}
\end{remark}

\begin{remark}
\label{Rem:11}
From the explicit form of matrix entries~\eqref{eq:matB} of the Gram matrix $\widehat{B}^{(p-1)}$, we see it is a block-diagonal matrix with respect to the upper pairs of the sub-indices $\left( \mu,\nu \right) $
and $\left( \mu ^{\prime },\nu ^{\prime }\right) $.  Non-zero elements are obtained  only for diagonal blocks satisfying $\left( \mu ,\nu \right) =\left(
\mu ^{\prime },\nu ^{\prime }\right) $, and they are of the form
\begin{equation}
\widehat{B}_{\tiny{\left[ 
\begin{array}{cc}
\mu  & \nu  \\ 
J_{\alpha } & K_{\beta }
\end{array}
,\omega \right] ,\left[ 
\begin{array}{cc}
\mu & \nu \\ 
I_{\alpha ^{\prime }} & L_{\beta ^{\prime }}
\end{array}
,\gamma \right] }}^{(p-1)}=\frac{1}{%
d^{p}(d^{2}-1)}\delta _{J_{\alpha }I_{\alpha ^{\prime }}}\delta _{K_{\beta
}L_{\beta ^{\prime }}}\left[ d\frac{m_{\mu }m_{\nu }}{m_{\omega }}\delta
^{\omega \gamma }-m_{\mu }\delta ^{\mu \nu }\right] ,
\end{equation}%
where $\alpha ,\alpha ^{\prime }\vdash \mu $ and $\beta ,\beta ^{\prime
}\vdash \nu .$
\end{remark}
Remark~\ref{Rem:11} allows us to investigate the interior structure of the blocks of the matrix $\widehat{B}^{(p-1)}$.

\begin{proposition}
\label{Prop:12}
The block-diagonal matrices $\widehat{B}^{(p-1)}_{\mu\nu }$, whose entries are defined in Remark~\ref{Rem:11}
\begin{equation}
\widehat{B}^{(p-1)}_{\mu\nu}=\left( \widehat{B}_{\tiny{\left[ 
\begin{array}{cc}
\mu  & \nu  \\ 
J_{\alpha } & K_{\beta }
\end{array}
,\omega \right] ,\left[ 
\begin{array}{cc}
\mu & \nu \\ 
I_{\alpha ^{\prime }} & L_{\beta ^{\prime }}
\end{array}
,\gamma \right] }}^{(p-1)}\right)
\end{equation}
have the following block-diagonal form 
\begin{equation}
\widehat{B}^{(p-1)}_{\mu\nu}=\mathbf{1}_{d_{\mu }d_{\nu }}\otimes B^{(p-1)}_{\mu\nu},
\end{equation}
where the each block $B^{(p-1)}_{\mu\nu}$ has matrix entries of the form
\begin{equation}
B^{(p-1)}_{\mu\nu}=(b_{\omega \gamma }^{\mu \nu })=\left(\frac{1}{d^{p}(d^{2}-1)}\left( d\frac{m_{\mu
}m_{\nu }}{m_{\omega }}\delta ^{\omega \gamma }-m_{\mu }\delta ^{\mu \nu
}\right)\right) \in M(n,\mathbb{R}).
\end{equation}
The parameter $n$ is the number of common irreps $\omega,\gamma$ of $\mathfrak{S}_{p-1}$ in irreps $\mu,\nu$ of $\mathfrak{S}_p$.
\end{proposition}

From the above proposition, we can deduce further properties of the block-diagonal matrices $\widehat{B}^{(p-1)}_{\mu\nu }$.

\begin{corollary}
The block-diagonal matrices $\widehat{B}^{(p-1)}_{\mu\nu }$ from Proposition~\ref{Prop:12}, have the following properties:
\begin{enumerate}[a)]
\item If $\mu \neq \nu $ and $\mu \sim _{\square }\nu$, then $B_{\mu \nu }^{(p-1)}=\frac{1}{d^{p}(d^{2}-1)}\left( d\frac{m_{\mu
}m_{\nu }}{m_{\omega }}\delta ^{\omega \gamma }-m_{\mu }\delta ^{\mu \nu
}\right)$ is one-dimensional and we have
\begin{equation}
\widehat{B}_{\mu \nu }^{(p-1)}=\mathbf{1}_{d_{\mu }d_{\nu }}\otimes B_{\mu \nu
}^{(p-1)}=\mathbf{1}_{d_{\mu }d_{\nu }} \otimes   \frac{1}{d^{p-1}(d^{2}-1)}
\frac{m_{\mu }m_{\nu }}{m_{\omega }},
\end{equation}
where $\omega =\mu -\Box \ \wedge \omega=\nu-\Box$.

\item If $\mu =\nu $, then 
\begin{equation}
\widehat{B}_{\mu \mu }^{(p-1)}=\mathbf{1}_{d_{\mu }^{2}}\otimes \frac{m_{\mu }}{
d^{p}(d^{2}-1)}\left( d\frac{m_{\mu }}{m_{\omega }}\delta ^{\omega \gamma
}-1\right) =\mathbf{1}_{d_{\mu }^{2}}\otimes B_{\mu \mu }^{(p-1)},
\end{equation}
where $\omega ,\gamma \in \widehat{\mathfrak{S}}_{p-1}$, and $\omega,\gamma=\mu-\Box$. Dimension of the matrix $
B_{\mu \mu }^{(p-1)}$ is equal to the number of irreps of $\mathfrak{S}_{p-1}$
in  the restriction of $\mu \in \widehat{\mathfrak{S}}_{p}$ to $\mathfrak{S}_{p-1}$.
\end{enumerate}
\end{corollary}

Let us summarize the above findings. For this reason let us introduce $\mu _{i}:i=1,2,\ldots,k$, which are all irreps of $\mathfrak{S}_{p}$ included in $V[\mathfrak{S}_{p}]$, i.e. for which the height $h(\mu_i)$ of the first column of $\mu_i$ satisfies $h(\mu_i)\leq d$.  Then we have the following alternative form of Corollary 32 contained in~\cite{studziński2025WBA}:

\begin{proposition}
The Gram matrix $\widehat{B}^{(p-1)}$ from Proposition~\ref{prop:multiB} that defines multiplication~\eqref{eq:compB} of the spanning operators $\widehat{G}_{\Gamma \Delta }^{(p-1)}$ in ideal $\widehat{\mathcal{M}}^{(p-1)}$, has
the following block diagonal form
\begin{equation}
\widehat{B}^{(p-1)}=\operatorname{diag}\left( \widehat{B}_{\mu _{1}\mu _{1}}^{(p-1)},\widehat{B}_{\mu
_{2}\mu _{2}}^{(p-1)},\ldots,\widehat{B}_{\mu _{k}\mu _{k}}^{(p-1)},\widehat{B}_{\mu _{j_{1}}\mu
_{j_{2}}}^{(p-1)},\widehat{B}_{\mu _{j_{1}}\mu _{j_{3}}}^{(p-1)},\ldots,\widehat{B}_{\mu
_{j_{l-1}}\mu _{j_{l}}}^{(p-1)}\right) ,
\end{equation}
where we have in $\widehat{B}_{\mu _{j_{a}}\mu _{j_{b}}}^{(p-1)}$ 
\begin{equation}
\mu _{j_{a}},\mu _{j_{b}}:a\neq b\quad \wedge \quad \mu _{j_{a}}\sim
_{\square }\mu _{j_{b}}.
\end{equation}%
This Gram matrix $\widehat{B}^{(p-1)}$ is invertible, if and only if 
\begin{equation}
\forall \mu _{i}:i=1,2,\ldots,k\quad dm_{\mu _{i}}\neq \sum_{\alpha _{k}=
\mu _{i}-\Box}m_{\alpha _{k}}.
\end{equation}
\end{proposition}

Using the results derived in Appendix B of~\cite{Moz1}, we present result concerning under what conditions generating operators  $\widehat{G}_{\Gamma \Delta }^{(p-1)}$ in ideal $\widehat{\mathcal{M}}^{(p-1)}$ form a quasi-matrix basis. Namely, we have:

\begin{theorem}
Non-trivial, generating operators $\widehat{G}_{\Gamma \Delta }^{(p-1)}$ of the ideal $\widehat{\mathcal{M}}^{(p-1)}$ from Definition~\ref{Not:1}, which are of the form  
\begin{equation}
\widehat{G}_{\tiny{\left[ 
\begin{array}{cc}
\mu _{1} & \mu _{2} \\ 
I_{\gamma } & K_{\eta }
\end{array}
,\kappa \right] ,\left[ 
\begin{array}{cc}
\mu _{3} & \mu _{4} \\ 
J_{\alpha } & L_{\beta }
\end{array}
,\tau \right]}}^{(p-1)}=E_{I_{\gamma }A_{\kappa }}^{\mu _{1}}\otimes
E_{K_{\eta }A_{\kappa }}^{\mu _{2}}Q^{(p-1)}E_{B_{\tau }J_{\alpha
}}^{\mu _{3}}\otimes E_{B_{\tau }L_{\beta }}^{\mu_{4}},
\end{equation}
where 
\begin{equation}
\left( \mu _{1}\neq \mu _{2}\quad \Rightarrow \quad \mu _{1}\sim _{\square
}\mu _{2}\right) \qquad \vee \qquad \left( \mu _{3}\neq \mu _{4}\quad
\Rightarrow \quad \mu _{3}\sim _{\square }\mu _{4}\right) 
\end{equation}
form  a quasi-matrix basis in the ideal $\widehat{\mathcal{M}}^{(p-1)}$,  if and only if the Gram matrix $\widehat{B}^{(p-1)}$ from Proposition~\ref{prop:multiB} is invertible, i.e. when
\begin{equation}
\forall \mu _{i}:i=1,2,\ldots,k\quad dm_{\mu _{i}}\neq \sum_{\alpha _{k}=
\mu _{i}-\Box}m_{\alpha _{k}}.
\end{equation}
\end{theorem}

Further, by  applying the general method of constructing pure matrix bases, presented in Appendix B of~\cite{Moz1}, one can obtain the following theorem:

\begin{theorem}
\label{thm:pureMatBas}
If the Gram matrix $\widehat{B}^{(p-1)}$ Proposition~\ref{prop:multiB} is invertible, then the
pure matrix basis generating the ideal $\widehat{\mathcal{M}}^{(p-1)}$ is given by the following
operators 
\begin{equation}
\label{Gwithouttilde}
G_{\Delta \Pi }^{(p-1)}=\sum_{\Lambda }\left( \widehat{B}^{(p-1)}\right)
_{\Lambda \Delta }^{-1}\widehat{G}_{\Lambda \Pi }^{(p-1)},
\end{equation}
which satisfy the matrix multiplication rule
\begin{equation}
\label{eq:multi1}
G_{\Gamma \Delta }^{(p-1)}G_{\Lambda \Pi }^{(p-1)}=\delta _{\Delta \Lambda
}G_{\Upsilon \Pi }^{(p-1)}.
\end{equation}
\end{theorem}

Thanks to the above theorem, we see that the new redefined operators~\eqref{Gwithouttilde}, spanning  the ideal $\widetilde{\mathcal{M}}^{(p-1)}$, form an orthonormal set of operators.

\section{Examples of irreducible matrix units construction for $\mathcal{A}^d_{2,2}$}
\label{sec:Example2v2}
In this case, for $p=2$, the operators $E^{\mu}_{ij}\in\mathbb{C}[\mathfrak{S}_2]$ are just the symmetric projector or the antisymmetric projector, and they are one-dimensional, with respective multiplicities equal to
\begin{equation}
m_{S}=\frac{1}{2}d(d+1),\quad m_{A}=\frac{1}{2}d(d-1),
\end{equation}
where $S,A$ denote symmetric/antisymmetric irrep of $\mathfrak{S}_2$. The operators $E^{\mu}_{ij}\in\mathbb{C}[\mathfrak{S}_2]$ will be denoted as $E_S, E_A$ for short. Additionally, we introduce  the trivial irrep $\operatorname{id}$ included in restrictions of irreps $A$ and $S$ to $\mathfrak{S}_1$, with $m_{\operatorname{id}}=d$.

According to Corollary~\ref{cor:1}, the algebra $\mathcal{A}_{2,2}^d$ is generated in the
following way 
\begin{align}
\label{eq:Alg_span}
\mathcal{A}_{2,2}^d=\operatorname{span}_{
\mathbb{C}}\{&\overrightarrow{E}_{i}\otimes \overleftarrow{E}_{ j}Q^{(2)}\overrightarrow{E}_{k}\otimes \overleftarrow{E}_{l}, \overrightarrow{E}_{i}\otimes \overleftarrow{E}_{j}Q^{(1)}\overrightarrow{E}_{k}\otimes \overleftarrow{E}_{l},\overrightarrow{E}_{i}\otimes \overleftarrow{E}_{j}Q^{(0)}\overrightarrow{E}_{k}\otimes \overleftarrow{E}_{l}^\},
\end{align}
where $i,j,k,l\in \{A,S\}$. 
As we can see from~\eqref{eq:Alg_span}, the algebra $\mathcal{A}_{2,2}^d$ is a sum (not direct) of three linear subspaces. For further simplification in this section, we omit the arrows above the operators $E_i$, remembering that they are constructed in opposite directions on both sides of the wall. 

Having the above introduction, we can proceed with the construction of the irreducible matrix units from Proposition~\ref{Prop:7}, producing the decomposition  $\mathcal{A}_{2,2}^d=\widetilde{\mathcal{M}}^{(2)} \oplus \widetilde{\mathcal{M}}^{(1)} \oplus \widetilde{\mathcal{M}}^{(0)}$. This can be achieved in three steps:
\begin{enumerate}[a)]
    \item First we consider the maximal ideal $\widetilde{\mathcal{M}}^{(2)}$. In this case, the matrix basis
in $\widetilde{\mathcal{M}}^{(2)}$ has the following form
\begin{equation}
\label{eq:BazaM2}
G_{kl}^{(2)}=\frac{1}{\sqrt{m_{k}m_{l}}}E_{k}\otimes \mathbf{1}Q^{(2)}E_{l}\otimes \mathbf{1},\quad G_{ij}^{(2)}G_{kl}^{(2)}=\delta _{jk}G_{il}^{(2)},
\end{equation}
where $k,l\in \{A,S,\}$. Note that this matrix algebra is two-dimensional independently of $d$.

Using equation~\eqref{eq:BazaM2}, we can formulate the following:

\begin{proposition}
\label{prop:projM2}
Projector $G^{(2)}$ onto ideal $\widetilde{\mathcal{M}}^{(2)}$ is of the form 
\begin{equation}
\label{eq:projM2}
G^{(2)}=\sum_{k\in \{A,S\}}G_{kk}^{(2)}=\sum_{k\in\{A,S\}}\frac{1}{m_{k}}E_{k}\otimes \mathbf{1}Q^{(2)}E_{k}\otimes \mathbf{1}.
\end{equation}
\end{proposition}

Operators from~\eqref{eq:BazaM2}, projector from~\eqref{eq:projM2}, and operator $Q^{(2)}$ from~\eqref{eq:Qs}, have respective traces equal to: 
\begin{equation}
\operatorname{Tr}(G^{(2)})=2,\quad \operatorname{Tr}(G_{AA}^{(2)})=1,\quad \operatorname{Tr}(G_{SS}^{(2)})=1,\quad \operatorname{Tr}(Q^{(2)})=1.
\end{equation}
From the above properties, we are in a position to formulate two facts describing further properties of the ideal $\widetilde{\mathcal{M}}^{(2)}$ and the operator $Q^{(2)}$.

\begin{fact}
The ideal $\widetilde{\mathcal{M}}^{(2)}$ satisfies the following property:
\begin{equation}
\widetilde{\mathcal{M}}^{(2)}=\operatorname{span}_{\mathbb{C}}\{E_i\otimes E_j Q^{(2)}E_k\otimes E_l\}=G^{(2)}\left( \operatorname{span}_{\mathbb{C}}\{E_i\otimes E_j Q^{(2)}E_k\otimes E_l\}\right) G^{(2)},
\end{equation}
where $i,j,k,l\in \{A,S\}$, operator $Q^{(2)}$ is diven in Definition~\ref{def:Q}, and $G^{(2)}$ is given through Proposition~\ref{prop:projM2}. In this case, the subspace $\operatorname{span}_{\mathbb{C}}\{E_i\otimes E_jQ^{(2)}E_k \otimes E_j\}\subset \mathcal{A}_{2,2}^d$ is in fact an ideal.
\end{fact}

The operators $G_{kl}^{(2)}$ from~\eqref{eq:BazaM2} form the pure matrix basis generating the ideal $\widetilde{\mathcal{M}}^{(2)}$, so they span the irrep, which is equal to the considered ideal~\cite{Littlewood}. Now, having the projector $G^{(2)}$ from Proposition~\ref{prop:projM2}, we can deduce that $\operatorname{Tr}(G^{(2)})=m_{\widetilde{\mathcal{M}}^{(2)}} d_{\widetilde{\mathcal{M}}^{(2)}}$. The numbers $m_{\widetilde{\mathcal{M}}^{(2)}}, d_{\widetilde{\mathcal{M}}^{(2)}}$ are multiplicity and dimension of the irrep $\widetilde{\mathcal{M}}^{(2)}$ in the space $(\mathbb{C}^{d})^{\otimes 2p}$ respectively. This allows us to deduce:

\begin{fact}
\label{fact:19}
 The multiplicity $m_{\widetilde{\mathcal{M}}^{(2)}}$ and the dimension $d_{\widetilde{\mathcal{M}}^{(2)}}$ in the space $(\mathbb{C}^{d})^{\otimes 2p}$ of the irrep of the algebra $\mathcal{A}^d_{2,2}$, defined by ideal $\widetilde{\mathcal{M}}^{(2)}$, are $m_{\widetilde{\mathcal{M}}^{(2)}}=1, d_{\widetilde{\mathcal{M}}^{(2)}}=2$. These numbers are independent of $d$.   
\end{fact}

Finally, we calculate matrix form of $Q^{(2)}$ in the irrep $\widetilde{\mathcal{M}}^{(2)}$.

\begin{proposition}
In the irrep $\widetilde{\mathcal{M}}^{(2)}$, the operator $Q^{(2)}$ is represented by the following matrix 
\begin{equation}
\varphi^{(2)}(Q^{(2)})=\frac{1}{d^{2}}\left( 
\begin{array}{cc}
m_{S} & \sqrt{m_{S}m_{A}} \\ 
\sqrt{m_{A}m_{S}} & m_{A}
\end{array}
\right) =\frac{1}{2d}\left( 
\begin{array}{cc}
d+1 & \sqrt{d^{2}-1} \\ 
\sqrt{d^{2}-1} & d-1
\end{array}
\right).
\end{equation}
\end{proposition}

\item Next, we consider the ideal $\widetilde{\mathcal{M}}^{(1)}$. Applying result of Proposition~\ref{Prop:7}, the non-trivial generators of this ideal are 
\begin{equation}
\widehat{G}_{[ij][kl]}^{(1)}=E_{i}\otimes E_{j}Q^{(1)}E_{k}\otimes E_{l},\quad
i,j,k,l\in \{A,S\},
\end{equation}
which, according to Proposition~\ref{prop:multiB}, satisfy
\begin{equation}
\widehat{G}_{[ab][ij]}^{(1)}\widehat{G}_{[kl][pq]}^{(1)}=\widehat{B}_{\left[
ij\right] \left[ kl\right] }^{(1)}\widehat{G}_{[ab][pq]}^{(1)},
\end{equation}
where
\begin{equation}
\label{eq:MMMM}
\widehat{B}^{(1)}=\left( \widehat{B}_{\left[ ij\right] \left[ kl\right]
}^{(1)}\right) =\frac{1}{4d}\left( 
\begin{array}{cccc}
d+2 &  &  &  \\ 
& d &  &  \\ 
&  & d &  \\ 
&  &  & d-2
\end{array}
\right),
\end{equation}
where we have the rows/columns of the above matrix are labelled in the following order $\left[ S,S\right] ,\left[ S,A\right], $ $\left[ A,S\right],\left[ A,A \right]$.

Invertibility of the matrix $\widehat{B}^{(1)}$ from~\eqref{eq:MMMM}, implies the possibility of constructing the pure matrix basis generating the ideal $\widetilde{\mathcal{M}}^{(1)}$, as it is described in Theorem~\ref{thm:pureMatBas}. In the notation adapted to this section, we would like to describe conditions under which the matrix $\widehat{B}^{(1)}$ can be inverted, and then construct the following set of pure matrix basis operators: 
\begin{equation}
G_{[kl][pq]}^{(1)}=\sum_{[rs]}\left( \widehat{B}^{(1)}\right) _{[rs][kl]}^{-1}
\widehat{G}_{[rs][pq]}^{(1)}.
\end{equation}

We summarize our results in the following proposition.

\begin{proposition}
For the matrix $\widehat{B}^{(1)}$ from~\eqref{eq:MMMM}, we have:
\begin{enumerate}[1)]
\item  If $d>2$ then matrix $\widehat{B}^{(1)}$ is invertible and the matrix
basis in the ideal $\widetilde{\mathcal{M}}^{(1)}$ can be written as 
\begin{align}
&G_{[SS][kl]}^{(1)}=\frac{4d}{d+2}\widehat{G}_{[SS][kl]}^{(1)}=\frac{4d}{d+2}
E_{S}\otimes E_{S}Q^{(1)}E_{k}\otimes E_{l},\quad k,l\in \{A,S\},\\
&G_{[SA][kl]}^{(1)}=4\widehat{G}_{[SA][kl]}^{(1)}=4E_{S}\otimes
E_{A}Q^{(1)}E_{k}\otimes E_{l},\quad k,l\in \{A,S\},\\
&G_{[AS][kl]}^{(1)}=4\widehat{G}_{[AS][kl]}^{(1)}=4E_{A}\otimes
E_{S}Q^{(1)}E_{k}\otimes E_{l},\quad k,l\in \{A,S\},\\
&G_{[AA][kl]}^{(1)}=\frac{4d}{d-2}\widehat{G}_{[AA][kl]}^{(1)}=\frac{4d}{d-2}
E_{A}\otimes E_{A}Q^{(1)}E_{k}\otimes E_{l},\quad k,l\in \{A,S\}.
\end{align}
The above operators satisfy the multiplication rule from~\eqref{eq:multi1} and form the four-dimensional matrix algebra. 

\item If $d=2$, then the matrix $\widehat{B}^{(1)}$ is not invertible, and we have
\begin{equation}
m_{A}=\frac{1}{2}d(d-1)=1,\quad m_{\operatorname{id}}=d=2.
\end{equation}
In this case, we have 
\begin{equation}
G_{[AA][kl]}^{(1)}=G_{[kl][AA]}^{(1)}=0,\quad k,l\in \{A,S\}.
\end{equation}%
The ideal $\widetilde{\mathcal{M}}^{(1)}$ is a three-dimensional irrep of the corresponding algebra with dimension $d_{\widetilde{\mathcal{M}}^{(1)}}=3$.
\end{enumerate}
\end{proposition}

We can go further with our consideration and construct the projector $G^{(1)}$ onto the ideal  $\widetilde{\mathcal{M}}^{(1)}$.

\begin{corollary}
\label{Cor:ProjM1}
The projector onto the ideal (irrep) $\widetilde{\mathcal{M}}^{(1)}$ can be written as
\begin{equation}
G^{(1)}=\sum_{k,l\in \{A,S\}}G_{[kl][kl]}^{(1)}=\sum_{k,l\in \{A,S\}}\left(\widehat{B}^{(1)}\right)_{\left[ kl
\right] \left[ kl\right] }^{-1}E_{k}\otimes E_{l}Q^{(1)}E_{k}\otimes E_{l},
\end{equation}
and 
\begin{equation}
\operatorname{Tr}(G^{(1)})=4(d^{2}-1),\quad \operatorname{Tr}(G_{[kl][kl]}^{(1)})=d^{2}-1,\quad
\operatorname{Tr}(Q^{(1)})=d^{2}-1.
\end{equation}
\end{corollary}

From the above properties, we are in a position to formulate two facts describing further properties of the ideal $\widetilde{\mathcal{M}}^{(1)}$ and the operator $Q^{(1)}$.

\begin{fact}
The ideal $\widetilde{\mathcal{M}}^{(1)}$ satisfies the following property:
\begin{equation}
\widetilde{\mathcal{M}}^{(1)}=G^{(1)}\left(\operatorname{span}_{\mathbb{C}}\{E_i\otimes E_jQ^{(1)}E_k\otimes E_l\}\right)G^{(1)}=\operatorname{span}_{
\mathbb{C}}\{E_i \otimes E_jQ^{(1)}E_k \otimes E_l\},
\end{equation}
where $i,j,k,l\in \{A,S\}$, operator $Q^{(1)}$ is given in Definition~\ref{def:Q}, and $G^{(1)}$ is given through Corollary~\ref{Cor:ProjM1}.
\end{fact}

Making a similar argumentation as was done for Fact~\ref{fact:19}, we formulate:

\begin{fact} 
The multiplicity $m_{\widetilde{\mathcal{M}}^{(1)}}$  of the irrep of the algebra defined by ideal the $\widetilde{\mathcal{M}}^{(1)}$,  in the space $(\mathbb{C}^{d})^{\otimes 2p}$, is $m_{\widetilde{\mathcal{M}}^{(1)}}=d^{2}-1$, and it depends on $d$. The dimension $d_{\widetilde{\mathcal{M}}^{(1)}}$ of the corresponding irrep $\widetilde{\mathcal{M}}^{(1)}$ of the algebra is $4$.
\end{fact}

\item For ideal $\widetilde{\mathcal{M}}^{(0)}$ we formulate the following first result:

\begin{proposition}
The projector $G^{(0)}$ onto ideal $\widetilde{\mathcal{M}}^{(0)}$ is equal to
\begin{equation}
G^{(0)}=\mathbf{1}-G^{(1)}-G^{(2)},
\end{equation}
where $G^{(2)}, G^{(1)}$ are projectors onto ideals $\widetilde{\mathcal{M}}^{(2)},\widetilde{\mathcal{M}}^{(1)}$ respectively given in Proposition~\ref{prop:projM2} and Corollary~\ref{Cor:ProjM1}.  
\end{proposition}

In the following, we show that the ideal $\widetilde{\mathcal{M}}^{(0)}$ is significantly more complicated than ideals $\widetilde{\mathcal{M}}^{(2)},\widetilde{\mathcal{M}}^{(1)}$ discussed before. We have

\begin{proposition}
The subspace $\mathcal{N}^{(0)}= \operatorname{span}_{\mathbb{C}}\{E_i \otimes E_jQ^{(0)}E_k \otimes E_l\}\subset \mathcal{A}_{2,2}^d$, where $i,j,k,l\in \{A,S\}$, and $Q^{(0)}$ is given Definition~\ref{def:Q},  is not an ideal in the algebra $\mathcal{A}_{2,2}^d$.
\end{proposition}

Indeed, for any $i,j\in \{A,S\}$, we have
\begin{equation}
Q^{(0)}E_i\otimes E_j Q^{(2)}=\delta
_{ij}\left( E_i\otimes \mathbf{1}-\frac{1}{2d}(d+j )\mathbf{1}\otimes \mathbf{1}\right) Q^{(2)},
\end{equation}
\begin{equation}
Q^{(0)}E_i\otimes E_jQ^{(1)}=\left(
E_i\otimes E_j-\frac{1}{d}(m_{i}m_{j}-m_{i}\delta_{ij})\mathbf{1} \otimes \mathbf{1}\right) Q^{(1)}.
\end{equation}
Which implies that 
\begin{equation}
\operatorname{span}_{\mathbb{C}}\{E_i\otimes E_j Q^{(0)}E_k \otimes E_l \}\widetilde{\mathcal{M}}^{(1)}\subset \widetilde{\mathcal{M}}^{(1)},\quad \operatorname{span}_{\mathbb{C}}\{E_i\otimes E_j Q^{(0)}E_k \otimes E_l \}\widetilde{\mathcal{M}}^{(2)}\subset \widetilde{\mathcal{M}}^{(2)}.
\end{equation}
Using the above argumentation and the identity resolution: 
\begin{equation}
\label{eq:Idres}
G^{(0)}=\mathbf{1}-G^{(1)}-G^{(2)}
\end{equation}
it follows

\begin{proposition}
The ideal $\widetilde{\mathcal{M}}^{(0)}$ has the following form 
\begin{equation}
\begin{split}
\widetilde{\mathcal{M}}^{(0)}=G^{(0)}\left(\operatorname{span}_{\mathbb{C}}\{E_i\otimes E_jQ^{(0)}E_k\otimes E_l\}\right)G^{(0)}=G^{(0)}\mathcal{N}^{0}G^{(0)}.
\end{split}
\end{equation}
It means that every element $x\in \widetilde{\mathcal{M}}^{(0)}$, due to~\eqref{eq:Idres}, can be decomposed as
\begin{equation}
\label{eq:xformM0}
x=n-G^{(1)}nG^{(1)}-G^{(2)}nG^{(2)},
\end{equation}
for some $n\in \mathcal{N}^{(0)}$.
\end{proposition}
Summarising, we see that in this case $\widetilde{\mathcal{M}}^{(0)}$ is not equal to $\mathcal{N}^{(0)}=\operatorname{span}_{\mathbb{C}}\{E_i\otimes E_j Q^{(0)}E_k\otimes E_l: i,j,k,l\in \{A,S\}\}$, as in the case of ideals $\widetilde{\mathcal{M}}^{(1)}$ and $\widetilde{\mathcal{M}}^{(2)}$. To construct a matrix basis for $\widetilde{\mathcal{M}}^{(0)}$, we first have to calculate the explicit form of an element $x \in \widetilde{\mathcal{M}}^{(0)}$, which is induced by~\eqref{eq:xformM0}.  After some long but straightforward calculations, we get

\begin{proposition}
An arbitrary element $n\in \mathcal{N}^{(0)}$ is of the form 
\begin{equation}
n=E_i\otimes E_jQ^{(0)}E_{k}\otimes E_l=\delta_{ik}\delta_{jl}E_i\otimes E_j-E_i\otimes E_jQ^{(1)}E_k\otimes E_l-E_i\otimes E_jQ^{(2)}E_k\otimes E_l
\end{equation}
and it generates element $x\in \widetilde{\mathcal{M}}^{(0)}$ of the form
\begin{equation}
x=\delta_{ik}\delta_{jl}\left( E_i\otimes E_j-(\widehat{B}^{(1)})^{-1}_{
\left[ ij\right] \left[ ij\right] }E_i\otimes E_j Q^{(1)}E_i\otimes E_j-\delta_{ij}\frac{1}{m_{i}}\mathbf{1}\otimes
E_iQ^{(2)}E_j\otimes \mathbf{1}\right) ,
\end{equation}
where $i,j,k,l \in \{A,S\}$.
\end{proposition}

The above proposition characterizes the $n \in \mathcal{N}^{(0)}$ that give rise to non-zero $G_{ij}^{(0)} \in \widetilde{\mathcal{M}}^{(0)}$, which in turn generate the ideal $\widetilde{\mathcal{M}}^{(0)}$.

\begin{corollary}
Only elements of the form
\begin{equation}
n=E_i\otimes E_jQ^{(0)}E_i\otimes E_j\in \mathcal{N}^{(0)}
\end{equation}
induce non-zero elements 
\begin{equation}
\label{eq:eq}
G_{ij}^{(0)}= E_i\otimes E_j-(\widehat{B}^{(1)})^{-1}_{\left[ ij\right] 
\left[ ij\right]}E_i\otimes E_jQ^{(1)}E_i\otimes E_j-\delta_{ij}\frac{1}{m_{i}}\mathbf{1}\otimes E_iQ^{(2)}E_j\otimes \mathbf{1} \in \widetilde{\mathcal{M}}^{(0)}.
\end{equation}
Equation~\eqref{eq:eq} implies that $G_{ij}^{(0)}$ are the generating elements for $\widetilde{\mathcal{M}}^{(0)}$
\begin{equation}
\begin{split}
\widetilde{\mathcal{M}}^{(0)}&=G^{(0)}\mathcal{N}^{(0)}G^{(0)}=\operatorname{span}_{\mathbb{C}}\{G_{ij}^{(0)}:i,j\in \{A,S\}\}.
\end{split}
\end{equation}
\end{corollary}

The next proposition gives the multiplication rule for the elements $G_{ij}^{(0)}$ and the decomposition of the ideal $\widetilde{\mathcal{M}}^{(0)}$ in a direct sum of smaller complex matrix ideals.

\begin{proposition}
Generators $G_{ij}^{(0)}:i,j\in \{A,S\}$ from~\eqref{eq:eq} satisfy the following multiplication rule
multiplication rule 
\begin{equation}
G_{ij}^{(0)}G_{kl}^{(0)}=\delta _{ik}\delta _{jl}G_{ij}^{(0)}.
\end{equation}
From this law it follows that the ideal $\widetilde{\mathcal{M}}^{(0)}$ is linearly generated by four orthogonal
projectors and 
\begin{equation}
\mathcal{\widetilde{M}}^{(0)}\cong\mathbb{C}\oplus \mathbb{C}\oplus \mathbb{C}\oplus \mathbb{C},
\end{equation}
is a direct sum of four one-dimensional matrix algebras.
\end{proposition}
\end{enumerate}

Now we investigate how the interior structure of the algebra $\mathcal{A}_{2,2}^d$ depends on the dimension parameter $d$.

\begin{proposition}
\begin{enumerate}[1)]
\item If $d>2$ then 
\begin{equation}
\operatorname{Tr}\left( G_{ij}^{(0)}\right) = 
\begin{cases}
 \frac{1}{4}(d^{2}-4)(d^{2}-1) \quad  \text{for} \quad  i\neq j,\\
 \frac{1}{4}d^{2}(d-1)(d+3) \quad \text{for} \quad i=j=S,\\ 
\frac{1}{4}d^{2}(d-3)(d+1) \quad \text{for} \quad i=j=A.
\end{cases}
\end{equation}
In the particular case of $d=3$, we have $G_{AA}^{(0)}=0$.

\item If $d=2$, then $G_{AA}^{(0)}=0$ and $G_{[AA][kl]}^{(1)}=G_{[kl][AA]}^{(1)}=0$ for  $k,l\in \{A,S\}$ (see above Prop. 22).
\end{enumerate}
\end{proposition}

We summarise our findings from this section and formulate our main result:
\begin{theorem}
\label{thm:Main}
The algebra $\mathcal{A}_{2,2}^d$ has the following matrix structure:
\begin{enumerate}
\item  If $d=2$, then 
\begin{equation}
\mathcal{A}_{2,2}^d=\widetilde{\mathcal{M}}^{(2)}\oplus \operatorname{span}_{\mathbb{C}}\{G_{[ij][kl]}^{(1)}:[ij],[kl]\neq \lbrack AA]\}\oplus \operatorname{span}_{\mathbb{C}}\{G_{ij}^{(0)}:(i,j)\neq (A,A)\}
\end{equation}
\begin{equation}
\mathcal{A}_{2,2}^d\cong M(2,\mathbb{C})\oplus M(3,\mathbb{C})\oplus \left( \mathbb{C}\oplus \mathbb{C}\oplus
\mathbb{C}\right) \Rightarrow \dim \mathcal{A}_{2,2}^d=16,
\end{equation}
where $M(m,\mathbb{C})$ is a complex matrix algebra of dimension $m$.

\item If $d=3$, then 
\begin{equation}
\mathcal{A}_{2,2}^d=\widetilde{\mathcal{M}}^{(2)}\oplus \operatorname{span}_{\mathbb{C}}\{G_{[ij][kl]}^{(1)}:i,j,k,l\in \{A,S\}\}\oplus \operatorname{span}_{\mathbb{C}}\{G_{ij}^{(0)}:(i,j)\neq (A,A)\}
\end{equation}
\begin{equation}
\mathcal{A}_{2,2}^d\cong M(2,\mathbb{C})\oplus M(4,\mathbb{C})\oplus \left( \mathbb{C}\oplus \mathbb{C}\oplus
\mathbb{C}\right) \Rightarrow \dim \mathcal{A}_{p=2}=23.
\end{equation}

\item If $d>3,$ then 
\begin{equation}
\mathcal{A}_{2,2}^d=\widetilde{\mathcal{M}}^{(2)}\oplus \operatorname{span}_{\mathbb{C}}\{G_{[ij][kl]}^{(1)}:i,j,k,l\in \{A,S\}\}\oplus \operatorname{span}_{\mathbb{C}}\{G_{ij}^{(0)}:i,j\in \{A,S\}\}
\end{equation}
\begin{equation}
\mathcal{A}_{2,2}^d\cong M(2,\mathbb{C})\oplus M(4,\mathbb{C})\oplus \left( \mathbb{C}\oplus \mathbb{C}\oplus
\mathbb{C}\oplus \mathbb{C}\right) \Rightarrow \dim \mathcal{A}_{2,2}^d=24. 
\end{equation}
\end{enumerate}
\end{theorem}

\section{Further properties of the algebra $\mathcal{A}^d_{p,p}$ and irreducible matrix units construction for $p=3$.}
\label{sec:Example3v3}

In this section, we continue investigating properties of the algebra $\mathcal{A}^d_{p,p}$. We start by formulating three results that hold for any value of $p$ and $d$. Then we fix $p=3$ and write explicitly the result of the squeezing of any $E_{ij}^\mu \otimes E_{kl}^\nu \in \mathbb{C}[\mathfrak{S}_3] \otimes \mathbb{C}[\mathfrak{S}_3]$ by the operator $V_{(3,3')}^{t_{3'}}$. We represent the result of the squeezing by the irreducible matrix units for the algebra $\mathcal{A}_{2,2}^d$, which are discussed in the previous section.  The squeezing lemmas play the central role in constructing respective irreducible matrix units in appropriate ideals of the considered algebra. For example, knowing the structure of squeezing by the element $V^{(p-1)}$ was necessary in constructing matrix units for the ideal $\mathcal{M}^{(p-1)}$ in~\cite{studziński2025WBA} or for the ideal $\widetilde{\mathcal{M}}^{(2)}$ in Section~\ref{sec:Example2v2}. We hope the result of such a kind can be used in further investigations in constructing the group-adapted irreducible matrix units.  

\subsection{Additional general considerations concerning the algebra $\mathcal{A}^d_{p,p}$}
Our considerations start from the following general fact for the single arc operators from the ideal $\mathcal{M}^{(1)}\subset \mathcal{M}^{(p)}$:

\begin{fact}
Let $a=1,2,\ldots ,p-1$ and $c'=1,2,\ldots,p'-1$, then
\begin{equation}
\begin{split}
&V^{t_{p'}}_{(p,p')}V_{(a,p)} V_{(p',c')}V^{t_{p'}}_{(p,p')}=V^{t_{p'}}_{(p,p')}V^{t_{c'} }_{(a,c')}=V_{(a,p-1)}V_{(p'-1,c')}V^{t_{p'}}_{(p,p')}V^{t_{p'-1}}_{(p-1,p'-1)}V_{(a,p-1)} V_{(p'-1,c')},
\end{split}
\end{equation}
and
\begin{equation}
V^{t_{p'}}_{(p,p')}V_{(a,p)}V^{t_{p'}}_{(p,p')}=V^{t_{p'}}_{(p,p')}=V^{t_{p'}}_{(p,p')}
V_{(p',c')}V^{t_{p'}}_{(p,p')}.
\end{equation}
\end{fact}

Next, we formulate a trace property for the operator $V^{t_{p'}}_{(p,p')}$, when acting on the irreducible matrix units $E_{ij}^\mu \otimes E_{kl}^\nu \in \mathbb{C}[\mathfrak{S}_p] \otimes \mathbb{C}[\mathfrak{S}_p]$:

\begin{proposition}
Let $V^{t_{p'}}_{(p,p')}$ be a swap operator $V_{(p,p')}\in \operatorname{End}[(\mathbb{C}^d)^{\otimes 2p}]$ between $p-$th and $p'-$th system, partially transposed over $p'-$th system. Then for any $p$ and $\mu ,\nu \in \widehat{\mathfrak{S}}_3$, we have
\begin{equation}
\operatorname{Tr}\left( V^{t_{p'}}_{(p,p')}E_{ij}^{\mu }\otimes E_{kl}^{\nu
}\right) =\delta _{ij}\delta _{kl}\frac{m_{\mu }m_{\nu }}{d}.
\end{equation}
\end{proposition}

We conclude our preparations for presenting the main results of this section by presenting the following general lemma concerning a certain decomposition of the irreducible matrix units $E_{ij}^\mu \in \mathbb{C}[\mathfrak{S}_p]$.

\begin{lemma}
Let $E_{I_{\gamma }J_{\alpha }}^{\mu }$ be irreducible matrix units for the group algebra $\mathbb{C}[\mathfrak{S}_p]$ written in notation from~\eqref{eq:notationPRIR2}. Then, for any irreps $\mu \in \widehat{\mathfrak{S}}_p$ and $\alpha ,\gamma = \mu- \Box,$ with the respective dimensions $d_\mu, d_\gamma$, we have
\begin{equation}
\begin{split}
E_{I_{\gamma }J_{\alpha }}^{\mu }&=\frac{d_{\mu }}{pd_{\gamma }}
\sum_{k_{\gamma }}E_{i_{\gamma }k_{\gamma }}^{\gamma }\left[
\sum_{a=1}^{p}\varphi _{J_{\alpha }K_{\gamma }}^{\mu }(a,p)V_{(a,p)}\right]\\
&=\frac{d_{\mu }}{pd_{\gamma }}
\sum_{k_{\gamma }}E_{i_{\gamma }k_{\gamma }}^{\gamma }\left[
\sum_{a=1}^{p-1}\varphi _{J_{\alpha }K_{\gamma }}^{\mu }(a,p)V_{(a,p)}\right] +\frac{
d_{\mu }}{pd_{\gamma }}\delta ^{\alpha \gamma }E_{i_{\gamma }j_{\alpha
}}^{\gamma },
\end{split}
\end{equation}
where $\varphi _{J_{\alpha }K_{\gamma }}^{\mu }(a,p)$ are matrix elements of the irrep $\mu \vdash p$ of $(a,p) \in \mathfrak{S}_p$ from~\eqref{eqn:basis_Eij} in the notation from ~\eqref{eq:notationPRIR1} and~\eqref{eq:notationPRIR2}.
The last term $E_{i_{\gamma }j_{\alpha
}}^{\gamma }$ of the above expression is defined outside of the support of $V^{t_{p'}}_{(p,p')}$.
\end{lemma}
The above statements play the central role in the next subsection when we present explicit construction of the irreducible matrix units for the algebra $\mathcal{A}^d_{3,3}$.  

\subsection{Construction of irreducible matrix units for $\mathcal{A}^d_{3,3}$}

From now on, we fix $p=3$, and work with the algebra $\mathcal{A}_{3,3}^d$. For simplification in this section, we omit the arrows above the operators $\overrightarrow{E}_{ij}^{\mu} \otimes \overleftarrow{E}^{\nu}_{kl}$ by writing just $E_{ij}^{\mu} \otimes E^{\nu}_{kl}$. The direction of basis construction will always be clear from the context. What is more, since we work with $\mathfrak{S}_3$, we have three irreps: $A-$ antisymmetric, $S-$ symmetric, and $C=(2,1)-$mixed symmetry. Projectors onto antisymmetric/symmetric irreps will be denoted as $E_A^{(3)},E_S^{(3)}$, respectively. We keep here the upper index indicating the number of systems, to underline the difference with the analogous projectors for $\mathfrak{S}_2$ irreps from the previous section. The irrep $C$ is 2-dimensional, and we denote the corresponding irreducible matrix units as $E^C_{ij} : i,j\in \{A,S\}$. The lower indices in this particular case, written in the basis reduced to $\mathfrak{S}_2$ group, can have only two values. This is because the diagram $C=(2,1)$, can be obtained by adding a single box $\Box$ to the symmetric/antisymmetric irreps of $\mathfrak{S}_2$. This notation keeps track of the mentioned box addition. Whenever we write $E_{ij}^\mu$, where $\mu \vdash 3$, we mean one of the operators from the set $\{E_A^{(3)}, E_S^{(3)}, E_{AA}^{C},E_{SS}^C, E_{AS}^C, E_{SA}^C\}.$

We start by formulating a statement on the squeezing of $E_{ij}^\mu \otimes E_{kl}^\nu \in \mathbb{C}[\mathfrak{S}_3] \otimes \mathbb{C}[\mathfrak{S}_3]$ by the single arc operator $V^{t_{3'}}_{(3,3')}$.

\begin{proposition}
\label{prop:squeezing_New}
Let $E_{ij}^\mu \otimes E_{kl}^\nu \in \mathbb{C}[\mathfrak{S}_3] \otimes \mathbb{C}[\mathfrak{S}_3]$, $V^{t_{3'}}_{(3,3')}$ be the single arc operator from the ideal $\mathcal{M}^{(1)}\subset \mathcal{A}_{3,3}^d$, and $G_{xy}^{(2)}, G_{[xy][qt]}^{(1)}, G_{xy}^{(0)}$ be the irreducible matrix units for the algebra $\mathcal{A}_{2,2}^d$ from Section~\ref{sec:Example2v2}. Then, for any $\mu ,\nu \in \widehat{\mathfrak{S}}_3$ we have decomposition:
\begin{equation}
\label{eq:squeezing_New}
V^{t_{3'}}_{(3,3')}E_{ij}^{\mu }\otimes E_{kl}^{\nu }V^{t_{3'}}_{(3,3')}=V^{t_{3'}}_{(3,3')}\otimes \left( \sum \lambda
_{xy}^{(2)}G_{xy}^{(2)}+\sum \lambda _{\lbrack xy][qt]}^{(1)}G_{[xy][qt]}^{(1)}+\sum
\lambda _{xy}^{(0)}G_{xy}^{(0)}\right),
\end{equation}
where $i,j,k,l,x,y,p,q,t\in \{A,S\}$, for $A,S \in \widehat{\mathfrak{S}}_2$, and $\lambda _{xy}^{(2)}, \lambda _{\lbrack xy][qt]}^{(1)}, \lambda _{xy}^{(0)} \in \mathbb{C}$. In the equation~\eqref{eq:squeezing_New}, we use compact notation for the coefficients in the decomposition, namely:
\begin{equation}
\lambda _{xy}^{(2)}\equiv \lambda _{xy}^{(2)}\left( 
\begin{array}{cc}
\mu & \nu \\ 
ij & kl
\end{array}
\right) ,\quad \lambda _{\lbrack xy][qt]}^{(1)}\equiv \lambda _{\lbrack
xy][qt]}^{(1)}\left( 
\begin{array}{cc}
\mu & \nu \\ 
ij & kl
\end{array}
\right) ,\quad \lambda _{xy}^{(0)}\equiv \lambda _{xy}^{(0)}\left( 
\begin{array}{cc}
\mu & \nu \\ 
ij & kl
\end{array}
\right).
\end{equation}
However, whenever the context is clear, we will use the simplified notation.
\end{proposition}

The next step is to derive the explicit form of the coefficients appearing in the right-hand side of decomposition~\eqref{eq:squeezing_New}. It is easy to see that:

\begin{align}
&\operatorname{Tr}\left( V^{t_{3'}}_{(3,3')}E_{ij}^{\mu }\otimes E_{kl}^{\nu
}V^{t_{3'}}_{(3,3')}G_{xy'}^{(2)}\right) =d\lambda
_{yx'}^{(2)},\\
&\operatorname{Tr}\left( V^{t_{3'}}_{(3,3')}E_{ij}^{\mu }\otimes E_{kl}^{\nu
}V^{t_{3'}}_{(3,3')}G_{[xy][uv]}^{(1)}\right) =d(d^{2}-1)\lambda
_{\lbrack uv][xy]}^{(1)},\\
&\operatorname{Tr}\left( V^{t_{3'}}_{(3,3')}E_{ij}^{\mu }\otimes E_{kl}^{\nu
}V^{t_{3'}}_{(3,3')}G_{xy}^{(0)}\right) =d\lambda _{yx}^{(0)}\operatorname{Tr}\left(
G_{xy}^{(0)}\right).
\end{align}

After involving calculations, we get expressions for the above coefficients. 
We summarize our calculations in the following proposition.
\begin{proposition}
The coefficients $\lambda_{yx'}^{(2)}, \lambda
_{\lbrack uv][xy]}^{(1)}, \lambda _{yx}^{(0)}$  from decomposition~\eqref{eq:squeezing_New} in Proposition~\ref{prop:squeezing_New}, are of the form:
\begin{enumerate}[1)]
    \item \begin{equation}
\lambda_{yx'}^{(2)}\equiv \lambda _{yx^{\prime }}^{(2)}\left( 
\begin{array}{cc}
\mu & \nu \\ 
ij & kl
\end{array}
\right) =\delta ^{\mu \nu }\delta _{jl}\delta _{ik}\delta _{lx}\delta _{iy}
\frac{1}{d^{2}}\frac{m_{\mu }}{\sqrt{m_{x}m_{y}}},
\end{equation}
which means that the non-zero coefficients are the following
\begin{equation}
\lambda _{ij'}^{(2)}\left( 
\begin{array}{cc}
\mu & \mu \\ 
ij & ij
\end{array}
\right) =\frac{1}{d^{2}}\frac{m_{\mu }}{\sqrt{m_{j}m_{i}}}.
\end{equation}

\item 
\begin{equation}
\lambda _{\lbrack xy][qt]}^{(1)}\equiv \lambda _{\lbrack xy][qt]}^{(1)}\left( 
\begin{array}{cc}
\mu  & \nu  \\ 
ij & kl
\end{array}
\right) =\frac{(\widehat{B}^{(1)})^{-1}_{[qt][qt]}}{d(d^{2}-1)}\delta _{xi}\delta _{yk}\delta
_{jq}\delta _{lt}\delta _{qt}\delta _{xy}\left( \delta _{xq}\frac{m_{\mu
}m_{\nu }}{m_{x}}-\frac{1}{d}\delta ^{\mu \nu }m_{\mu }\right) ,
\end{equation}
where $x\vdash \mu $ and $x\vdash \nu $ and the following coefficients
survive
\begin{equation}
\lambda _{\lbrack ii][jj]}^{(1)}\left( 
\begin{array}{cc}
\mu  & \nu  \\ 
ij & ij
\end{array}
\right) =\frac{(\widehat{B}^{(1)})^{-1}_{[jj][jj]}}{d(d^{2}-1)}\left( \delta _{ij}\frac{m_{\mu
}m_{\nu }}{m_{i}}-\frac{1}{d}\delta ^{\mu \nu }m_{\mu }\right) ,
\end{equation}
where the numbers $\widehat{B}^{(1)}_{[jj][jj]}=m_{j}(m_{j}-1)$.

\item 
\begin{equation}
\begin{split}
\lambda _{xy}^{(0)}\equiv \lambda _{xy}^{(0)}\left( 
\begin{array}{cc}
\mu  & \nu  \\ 
ij & kl
\end{array}
\right) &=\frac{1}{d\operatorname{Tr}(G_{xy}^{(0)})}\left( \delta _{xj}\delta _{ix}\delta
_{ly}\delta _{xy}m_{\mu }m_{\nu }\right) -\\
&-\frac{1}{d\operatorname{Tr}(G_{xy}^{(0)})}\left( \delta _{xi}\delta _{yk}\delta _{jx}\delta
_{ly}\delta _{xy}(\widehat{B}^{(1)})^{-1}_{[xy][xy]}\left[ \frac{m_{\mu }m_{\nu }}{m_{x}}-
\frac{1}{d}\delta ^{\mu \nu }m_{\mu }\right] \right) -\\
&-\frac{1}{d\operatorname{Tr}(G_{xy}^{(0)})}\left( \delta _{xy}\delta ^{\mu \nu }\delta
_{xl}\delta _{yi}\delta _{jx}\delta _{ky}\frac{1}{d}\frac{m_{\mu }}{\sqrt{
m_{x}m_{y}}}\right),
\end{split}
\end{equation}
and here, the surviving coefficients are 
\begin{equation}
\lambda _{ik}^{(0)}\left( 
\begin{array}{cc}
\mu  & \nu  \\ 
ii & kk
\end{array}
\right)
=\frac{1}{d\operatorname{Tr}(G_{xy}^{(0)})}\left( m_{\mu }m_{\nu }-\delta
_{ik}(\widehat{B}^{(1)})^{-1}_{[ik][ik]}\left[ \frac{m_{\mu }m_{\nu }}{m_{i}}-\frac{1}{d}
\delta ^{\mu \nu }m_{\mu }\right] -\delta _{ik}\delta ^{\mu \nu }\frac{1}{d}
\frac{m_{\mu }}{\sqrt{m_{i}m_{k}}}\right)
\end{equation}
and here 
\begin{equation}
\widehat{B}^{(1)}_{[ik][ik]}=
\begin{cases}
m_{i}m_{k}:i\neq k, \\ 
m_{i}(m_{i}-1):i=k.
\end{cases}
\end{equation}
\end{enumerate}
\end{proposition}

We summarize all the above findings by formulating the main result for this section.

\begin{theorem}
\label{thm:squeezingNew}
Let $E_{ij}^\mu \otimes E_{kl}^\nu \in \mathbb{C}[\mathfrak{S}_3] \otimes \mathbb{C}[\mathfrak{S}_3]$, and $V^{t_{3'}}_{(3,3')}$ be the single arc operator from the ideal $\mathcal{M}^{(1)}\subset \mathcal{A}_{3,3}^d$.  Then, for any $\mu ,\nu \in \widehat{\mathfrak{S}}_3$ we have decomposition:
\begin{equation}
\begin{split}
&V^{t_{3'}}_{(3,3')}E_{ij}^{\mu }\otimes E_{kl}^{\nu }V^{t_{3'}}_{(3,3')}= \delta _{ik}\delta _{jl}
\frac{1}{d}\frac{m_{\mu }}{\sqrt{m_{i}m_{j}}}G_{ij}^{(2)}V^{t_{3'}}_{(3,3')}+\\
&+\frac{(\widehat{B}^{(1)})^{-1}_{[ij][ij]}}{d(d^{2}-1)}
\left( \delta _{ij}\frac{m_{\mu }m_{\nu }}{m_{i}}-\frac{1}{d}\delta ^{\mu
\nu }m_{\mu }\right) G_{[ii][jj]}^{(1)}V^{t_{3'}}_{(3,3')}+\\
&+\frac{1}{d\operatorname{Tr}(G_{xy}^{(0)})}\left( m_{\mu
}m_{\nu }-\delta _{ik}(\widehat{B}^{(1)})^{-1}_{[ik][ik]}\left[ \frac{m_{\mu }m_{\nu }}{m_{i}}-
\frac{1}{d}\delta ^{\mu \nu }m_{\mu }\right] -\delta _{ik}\delta ^{\mu \nu }
\frac{1}{d}\frac{m_{\mu }}{\sqrt{m_{i}m_{k}}}\right) G_{ik}^{(0)}V^{t_{3'}}_{(3,3')},
\end{split}
\end{equation}
where
\begin{align}
&G_{kl}^{(2)}=\frac{1}{\sqrt{m_{k}m_{l}}}\mathbf{1}\otimes E_{k}Q^{(2)}E_{l}\otimes \mathbf{1},\quad G_{ij}^{(2)}G_{kl}^{(2)}=\delta _{jk}G_{il}^{(2)},\\
&G_{[ii][jj]}^{(1)}=(\widehat{B}^{(1)})^{-1}_{\left[ ii\right] \left[ ii\right]
}E_{i}\otimes E_{i}Q^{(1)}E_{j}\otimes E_{j},\\
&G_{ik}^{(0)}=E_{i}\otimes E_{k}-(\widehat{B}^{(1)})^{-1}_{\left[ ik\right] \left[ ik
\right] }E_{i}\otimes E_{k}Q^{(1)}E_{i}\otimes E_{k}-\delta
^{ik}\frac{1}{m_{i}}\mathbf{1}\otimes E_{i}Q^{(2)}E_{k}\otimes \mathbf{1}
\end{align}
are the group-adapted irreducible matrix units for the algebra $\mathcal{A}_{2,2}^d$ from Section~\ref{sec:Example2v2}.
\end{theorem}
The above theorem provides an explicit version of Lemma~\ref{eq:F1}, where the operator $X_{p\setminus r,p\setminus r}$ is expressed in terms of the group-adapted irreducible matrix units of the algebra $A_{2,2}^d$. This section illustrates the general difficulty of obtaining a refined decomposition of the contraction results computed within different ideals $\widetilde{\mathcal{M}}^{(r)}$.

\section{Discussion}
\label{sec:disscus}
In this work, we present an iterative construction of irreducible matrix units for the algebra of partially transposed permutation operators $\mathcal{A}^d_{p,p}$, which serves as a matrix representation for the diagrammatic walled Brauer algebra. Our approach, in contrast to earlier established Gelfand-Tsetlin type construction~\cite{grinko2023gelfandtsetlinbasispartiallytransposed}, relies on the following novel features:
\begin{enumerate}
    \item Obtained irreducible matrix units are group-adapted to the action of the algebra $\mathbb{C}[\mathfrak{S}_p] \times \mathbb{C}[\mathfrak{S}_p]$, where $\mathfrak{S}_p$ is the symmetric group. This substantially extends the previous results, where the basis with the group-adapted property has been constructed only in the highest and the second-highest ideals of the algebra. As an example, we present the full set of the matrix units for the case when $p=2$.
    \item The algorithm outputs a basis yielding a decomposition of $\mathcal{A}^d_{p,p}$ into a direct sum of ideals. The decomposition of this kind also strongly generalizes the previous approaches, where such a direct sum has been constructed only for the first two highest ideals~\cite{studziński2025WBA}. We explicitly present such a decomposition for the algebra $\mathcal{A}^d_{2,2}$.
\end{enumerate}
We hope to apply our algorithm to the area of higher-order quantum operations, which recently attracted some attention in the scientific community~\cite{taranto2025higherorderquantumoperations}. In particular, the presented algorithm can play a central role in so-called parallel configurations — for example, in transforming an unknown unitary program~\cite{yoshida2023reversing}, where symmetry of the type $\mathbb{C}[\mathfrak{S}_p] \times \mathbb{C}[\mathfrak{S}_p]$ arises naturally. In a more general context, this algorithm can potentially reduce the complexity of semi-definite programs with symmetries, similarly to~\cite{grinko2023linearprog}. The mentioned basis can also potentially be applied to construct multi-matrix invariants, and in particular, the scalar multi-trace operators, as explained in~\cite{Padellaro2025}.

We are aware of the weaknesses of our approach and the necessity of further research. To construct the irreducible matrix units in the $ r$-th step, one requires information from all previous steps. This is a substantial limitation when we increase the number of systems under consideration. However, the presented approach can be efficient when the number of systems is not too large and supported by a numerical approach. 

\section*{Acknowledgements}
MS is supported by the National Science Centre, Poland, Grant Sonata Bis 14 no. 2024/54/E/ST2/00316.
MH acknowledges support by the IRA Programme, project no. FENG.02.01-IP.05-0006/23, financed by the FENG program 2021-2027, Priority FENG.02, Measure FENG.02.01., with the support of the FNP.

\appendix
\section{Appendix}
\label{app:A}

In this appendix, we rewrite the statements of Lemma 7  and Theorem 9 from~\cite{studziński2025WBA} in a simpler form by the notation introduced around equations~\eqref{eq:notationPRIR1} and~\eqref{eq:notationPRIR2}.

\begin{lemma}
\label{lemWBA}
Let $E_{I_{\gamma }J_{\alpha' }}^{\mu }, E_{K_{\beta
}L_{\beta'}}^{\nu }$ be irreducible matrix units for the group algebra $\mathbb{C}[\mathfrak{S}_p]$ written in the notation from~\eqref{eq:notationPRIR2}, and let $V^{(p-1)}$ be an arc operator given through~\eqref{eq:Vl}, then: 
\begin{equation}
\operatorname{Tr}\left[ E_{I_{\alpha }J_{\alpha'}}^{\mu }\otimes E_{K_{\beta
}L_{\beta'}}^{\nu }V^{(p-1)}\right] =\frac{m_{\mu }m_{\nu }}{%
m_{\alpha }}\delta ^{\alpha \beta'}\delta _{I_{\alpha }K_{\beta
}}\delta _{J_{\alpha'}L_{\beta'}}.
\end{equation}
\end{lemma}

\bigskip

\begin{theorem}
\label{thmWBA}
Let $E_{I_{\gamma }J_{\alpha' }}^{\mu }, E_{K_{\beta
}L_{\beta'}}^{\nu }$ be irreducible matrix units for the group algebra $\mathbb{C}[\mathfrak{S}_p]$ written in the notation from~\eqref{eq:notationPRIR2}, and let $V^{(p)}, V^{(p-1)}$ be arc operators given through~\eqref{eq:Vl}, then: 
\begin{equation}
\label{eq:WBA_praca}
V^{(p-1)}E_{I_{\alpha }J_{\alpha'}}^{\mu }\otimes E_{K_{\beta
}L_{\beta'}}^{\nu }V^{(p-1)}=a_{\ I_{\alpha }J_{\alpha'}K_{\beta }L_{\beta}}^{\mu \ \qquad \nu }V^{(p)}+b_{\ I_{\alpha }J_{\alpha'}K_{\beta }L_{\beta'}}^{\mu
\ \qquad \nu }V^{(p-1)},
\end{equation}
where 
\begin{equation}
a_{\ I_{\alpha }J_{\alpha ^{\prime }}K_{\beta }L_{\beta ^{\prime }}}^{\mu \
\qquad \nu }=\frac{1}{d(d^{2}-1)}\left( dm_{\nu }\delta ^{\mu \nu }-\frac{%
m_{\mu }m_{\nu }}{m_{\alpha }}\delta ^{\alpha \alpha ^{\prime }}\right)
\delta _{I_{\alpha }K_{\beta }}\delta _{J_{\alpha ^{\prime }}L_{\beta
^{\prime }}},
\end{equation}%
\begin{equation}
b_{\ I_{\alpha }J_{\alpha ^{\prime }}K_{\beta }L_{\beta ^{\prime }}}^{\mu \
\qquad \nu }=\frac{1}{d(d^{2}-1)}\left( \frac{dm_{\mu }m_{\nu }}{m_{\alpha }}%
\delta ^{\alpha \alpha ^{\prime }}-m_{\nu }\delta ^{\mu \nu }\right) \delta
_{I_{\alpha }K_{\beta }}\delta _{J_{\alpha ^{\prime }}L_{\beta ^{\prime }}}.
\end{equation}
\end{theorem}

\printbibliography
\end{document}